\documentclass[%
reprint,
superscriptaddress,
preprintnumbers,
bibnotes,
amsmath,amssymb,
aps,
prl,
nolongbibliography,
]{revtex4-2}

\usepackage{graphicx}
\usepackage{dcolumn}
\usepackage{bm}
\usepackage{chemformula}
\usepackage{hyperref}

\begin{document}


\title{Spin-orbit phase behaviors of \ch{Na_2Co_2TeO_6} at low temperatures}

\author{Wenjie~Chen}
\thanks{These authors contributed equally to this study.}
\affiliation{International Center for Quantum Materials, School of Physics, Peking University, Beijing 100871, China}
\author{Xintong~Li}
\thanks{These authors contributed equally to this study.}
\affiliation{International Center for Quantum Materials, School of Physics, Peking University, Beijing 100871, China}
\author{Zhenhai~Hu}
\thanks{These authors contributed equally to this study.}
\affiliation{International Center for Quantum Materials, School of Physics, Peking University, Beijing 100871, China}
\author{Ze~Hu}
\affiliation{Renmin University of China, Beijing 100872, China}
\author{Li~Yue}
\affiliation{International Center for Quantum Materials, School of Physics, Peking University, Beijing 100871, China}
\author{Ronny~Sutarto}
\affiliation{Canadian Light Source, Saskatoon, Saskatchewan S7N 2V3, Canada}
\author{Feizhou~He}
\affiliation{Canadian Light Source, Saskatoon, Saskatchewan S7N 2V3, Canada}
\author{Kazuki~Iida}
\affiliation{Neutron Science and Technology Center, Comprehensive Research Organization for Science and Society (CROSS), Tokai, Ibaraki 319-1106, Japan}
\author{Kazuya~Kamazawa}
\affiliation{Neutron Science and Technology Center, Comprehensive Research Organization for Science and Society (CROSS), Tokai, Ibaraki 319-1106, Japan}
\author{Weiqiang~Yu}
\email{wqyu\_phy@ruc.edu.cn}
\affiliation{Renmin University of China, Beijing 100872, China}
\author{Xi~Lin}
\email{xilin@pku.edu.cn}
\affiliation{International Center for Quantum Materials, School of Physics, Peking University, Beijing 100871, China}
\affiliation{Beijing Academy of Quantum Information Sciences, Beijing 100193, China}
\affiliation{CAS Center for Excellence in Topological Quantum Computation, University of Chinese Academy of Sciences, Beijing 100190, China}
\author{Yuan~Li}
\email{yuan.li@pku.edu.cn}
\affiliation{International Center for Quantum Materials, School of Physics, Peking University, Beijing 100871, China}

\date{\today}

\begin{abstract}
We present a comprehensive study of single crystals of \ch{Na_2Co_2TeO_6}, a putative Kitaev honeycomb magnet, focusing on its low-temperature phase behaviors. A new thermal phase transition is identified at 31.0 K, below which the system develops a two-dimensional (2D) long-range magnetic order. This order precedes the well-known 3D order below 26.7 K, and is likely driven by strongly anisotropic interactions. Surprisingly, excitations from the 3D order do not support the order's commonly accepted ``zigzag'' nature, and are instead consistent with a ``triple-$\mathbf{q}$'' description. The 3D order exerts a fundamental feedback on high-energy excitations that likely involve orbital degrees of freedom, and it remains highly frustrated until a much lower temperature is reached. These findings render \ch{Na_2Co_2TeO_6} a spin-orbit entangled frustrated magnet that hosts very rich physics.

\end{abstract}

\maketitle



\begin{figure}[b]
\includegraphics[width=0.5\textwidth]{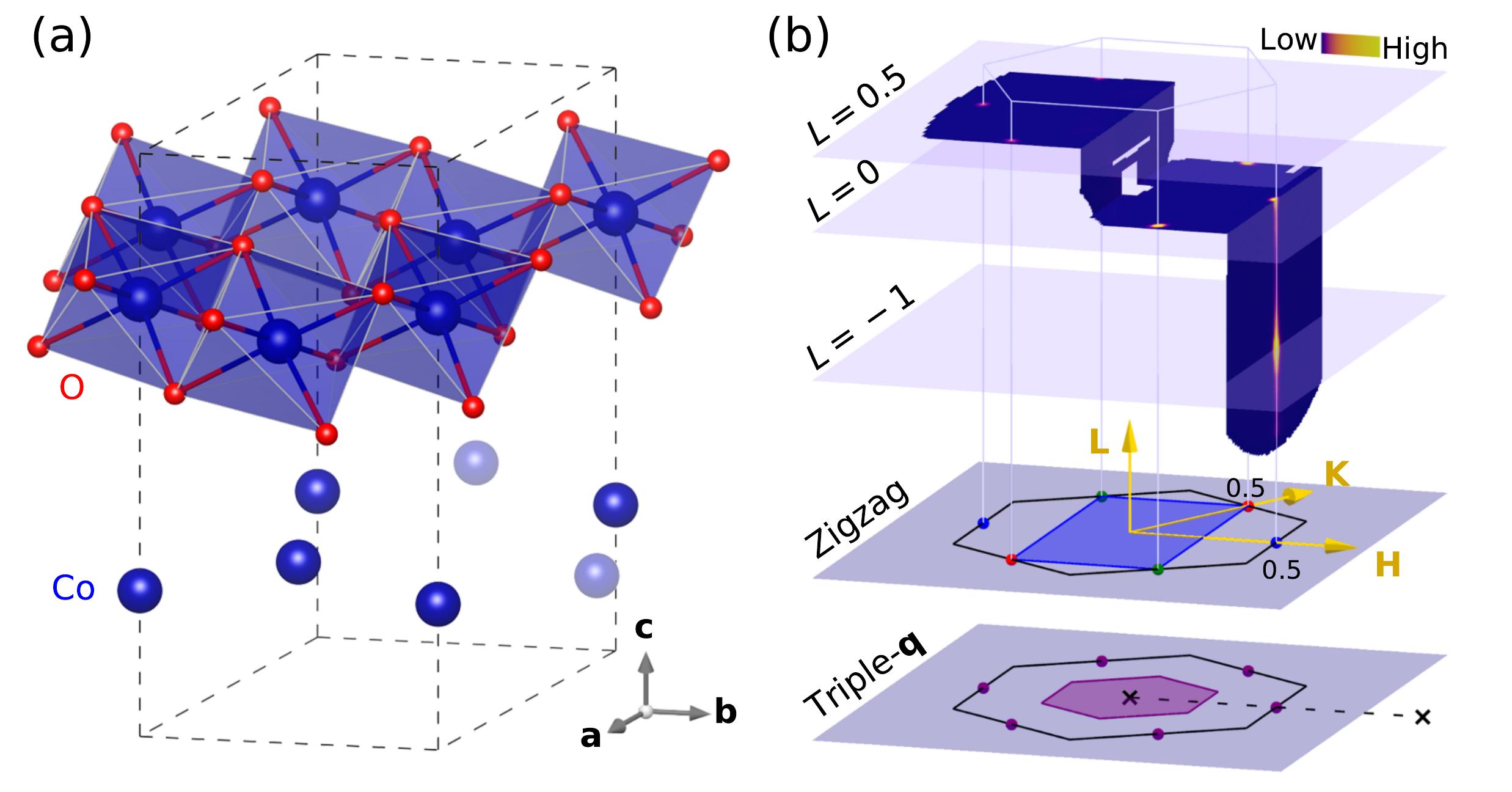}
\caption{\label{fig:1} (a) Co-O layers of \ch{Na_2Co_2TeO_6}. Co are at the centers of the edge-sharing O octahedra, forming a honeycomb lattice (lower layer, where O are not shown and Co outside the dashed unit cell are faded). (b) Upper half: momentum slices of neutron diffraction raw data obtained at $T = 5$ K. Sharp magnetic diffractions are seen at the $M$-points of the 2D structural Brillouin zone (BZ), consisting of rod-like signals running along $L$ from the 2D order, and peaks at integer $L$ (in reciprocal lattice units, used throughout this work) from the 3D order. Lower half: 2D structural (black hexagons) and magnetic BZs under the zigzag (blue rectangle, for domains characterized by the blue $M$-point) and triple-$\mathbf{q}$ (purple hexagon) schemes, see text. Colored spheres are $M$-points of the BZ. Dashed line illustrates the cut displayed in Fig.~\ref{fig:3}(a).}
\end{figure}

The exactly solvable Kitaev honeycomb model \cite{Kitaev06} has evoked considerable research interest in recent years, as it offers a distinct route to quantum spin liquids (QSLs). Realizing the model in real materials requires specific crystal structures and magnetic interactions \cite{Takagi19}. Former studies showed that bond-dependent Ising interactions, also known as Kitaev interactions, can arise from spin-orbit entangled pseudospin-1/2 degrees of freedom of $d^5$ transition metal ions situated in edge-shared octahedral crystal fields, and that a honeycomb lattice of such ions might approximate the Kitaev model \cite{Jackeli09}, as is believed to be the case for \ch{Na_2IrO_3} \cite{ChaloupkaPRL2010} and $\alpha$-\ch{RuCl_3} \cite{PlumbPRB2014}. More recent studies showed that $d^7$ ions Co$^{2+}$ with a high-spin $t_{2g}^5e_g^2$ configuration can also provide pseudospin-1/2 degrees of freedom with Kitaev interactions \cite{Liu18, Sano18}, as long as non-octahedral crystal fields are weak enough to leave the spin-orbit entanglement intact. But since Co has weaker spin-orbit coupling than Ir and Ru, this last hypothesis requires close scrutiny. Two layered cobaltates, \ch{Na_2Co_2TeO_6} and \ch{Na_3Co_2SbO_6}, in which oxygen ligands of Co$^{2+}$ form a nearly regular octahedron with small trigonal distortions \cite{ViciuJSSC2007}, have been proposed to potentially realize the Kitaev model \cite{Liu18,LiuPRL2020}. The Co-O sublattice of \ch{Na_2Co_2TeO_6} is displayed in Fig.~\ref{fig:1}(a).

\begin{figure*}[!ht]
\includegraphics[width=0.85\textwidth]{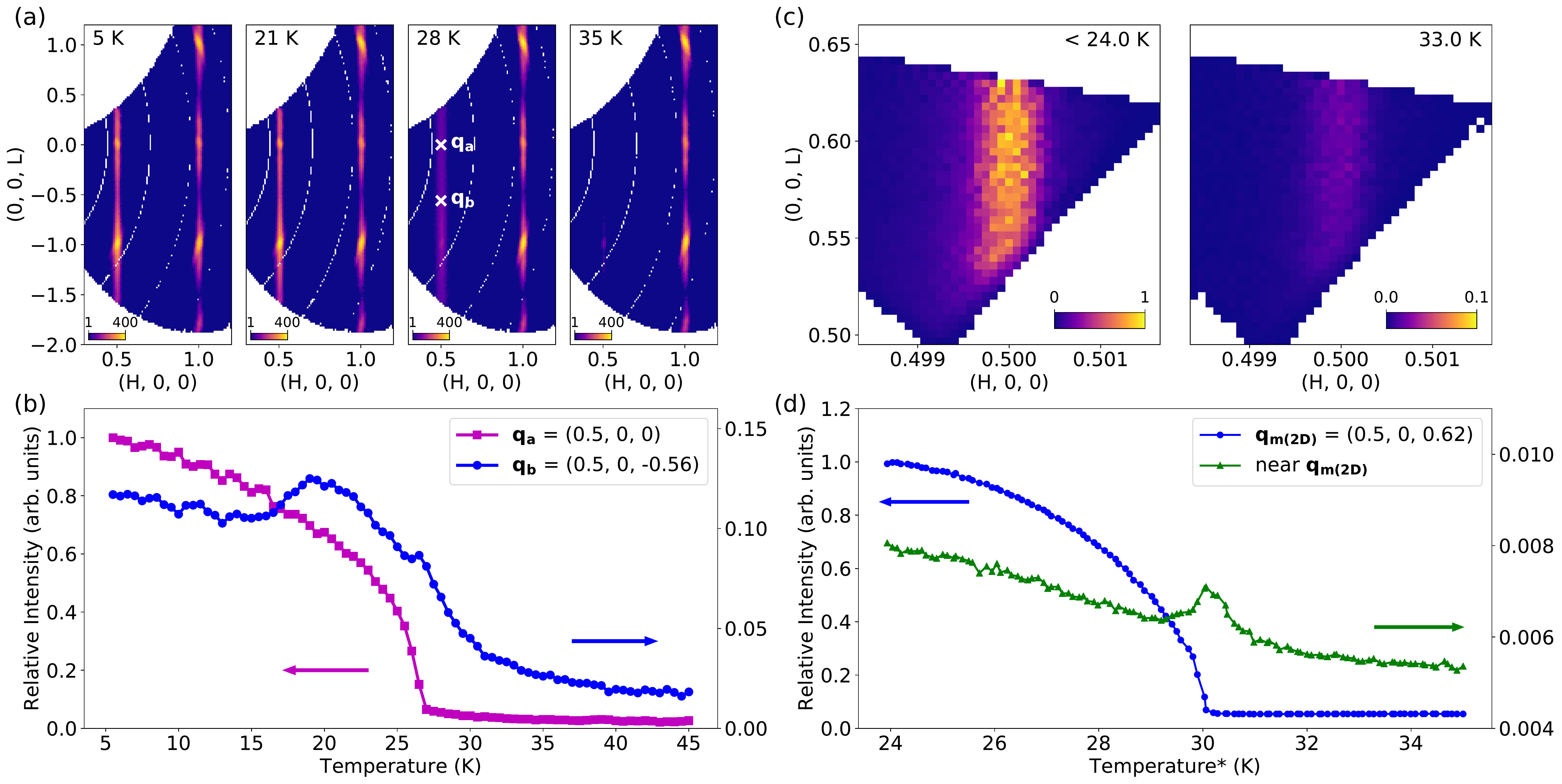}
\caption{\label{fig:2} (a) Neutron diffraction in the ($H$, 0, $L$) plane at different $T$. Intensities are plotted in false colors on a log scale. (b) $T$ dependence of signal at two $\mathbf{q}$ positions indicated by ``x'' in (a), characterizing 3D and 2D correlations (see text). (c) RXD in the ($H$, 0, $L$) plane. The signal is primarily due to magnetic scattering at $T < 24$~K, whereas the weak remaining intensity (note the different color scales) at $T=33$~K is due to a weak superstructure in the crystal lattice that likely originates from the Na layers (Fig.~S1 in \cite{SM}), which we consider unimportant here. (d) $T$ dependence of the RXD signals. Critical scattering (triangles) was measured in the same run by regions on the area detector surrounding the Bragg peak (circles) (Fig.~S2 in \cite{SM}). The temperature reading had not been calibrated at the sample position (see text).}
\end{figure*}

In most candidate Kitaev magnets, the systems develop long-range magnetic orders at low temperatures, instead of having a QSL ground state \cite{Takagi19}. Nevertheless, for understanding the materials and assessing their likelihood of realizing QSLs, such order may be instrumental because the symmetry of the order, as well as the associated magnon spectrum, may be used to determine the magnetic interactions \cite{BanerjeeScience2017,WinterNatCommun2017,RanPRL2017}. The information can be used to estimate whether and how the ground state can be tuned towards QSLs \cite{LiuPRL2020,WinterPRL2018}, and it provides the base for understanding exotic excitations \cite{BanerjeeScience2017,DoNatPhys2017}. The most commonly found long-range order is the so-called zigzag antiferromagnetic order, which is believed to be shared by the aforementioned Ir-, Ru-, and Co-based compounds \cite{LiuPRB2011,FengYe12, Choi12,Johnson15, Sears15,Lefrancois16, Bera17,WongJSSC2016,YanPRM2019}. Historically, observation of zigzag order had been considered promising for finding a nearby QSL phase, because the two phases are adjacent in the parameter space of the Kitaev-Heisenberg model \cite{ChaloupkaPRL2013}. However, this notion only applies to antiferromagnetic Kitaev interactions ($K>0$), yet all of the above systems are considered to have $K<0$ \cite{KatukuriNJP2014,YamajiPRL2014,SizyukPRB2014,HuPRL2015,
WinterPRB2016,KimPRB2016,RanPRL2017,Liu18,Sano18}, even though the Co-based cases are less certain according to recent studies \cite{Songvilay20,Lin20,Kim20}. When additional interactions (further-neighbor ones in particular) are included to rationalize the observation of zigzag order, theoretical phase diagrams become complicated \cite{KatukuriNJP2014,RauPRL2014,SizyukPRB2014,WinterPRB2016,KimchiPRB2011}, and it depends on model details whether a QSL phase remains accessible nearby. In this regard, the Co-based systems are likely dominated by nearest-neighbor interactions because their $3d$ orbitals are more localized than $4d$ and $5d$ \cite{LiuPRL2020}. While this may help simplify theories, it also raises the question about the stability of the zigzag order, which becomes particularly important for shaping our understanding of these $3d$ systems.

Here we present a systematic study of high-quality single crystals \cite{XiaoCGD2019,Yao20} of \ch{Na_2Co_2TeO_6}, bringing insight to the above two fronts. We show that the system's first thermal transition, out of the paramagnetic phase, is towards a hitherto unnoticed two-dimensional (2D) long-range magnetic order, which precedes the formation of the 3D order that is commonly considered of zigzag nature. As the Mermin-Wagner theorem precludes spontaneous breaking of continuous symmetry at finite temperature in 2D systems, the 2D order must hence break discrete symmetry, which would be natural if the order is driven by strongly anisotropic interactions such as $K$. We furthermore find a surprising result about the 3D order: magnons deeply in the ordered state are \textit{not} described by Brillouin zones (BZs) of zigzag order; instead, they seem to arise from a triple-$\mathbf{q}$ ground state, which may be indistinguishable from zigzag order in diffraction experiments performed on powder \cite{Lefrancois16, Bera17} or multi-domain samples. We see experimental signs that the 3D order is strongly frustrated upon its initial formation, and that it affects the high-energy orbital degrees of freedom. These results indicate that spin-orbit physics and frustration are at the heart of magnetism in \ch{Na_2Co_2TeO_6}.

\begin{figure*}[!ht]
\includegraphics[width=0.85\textwidth]{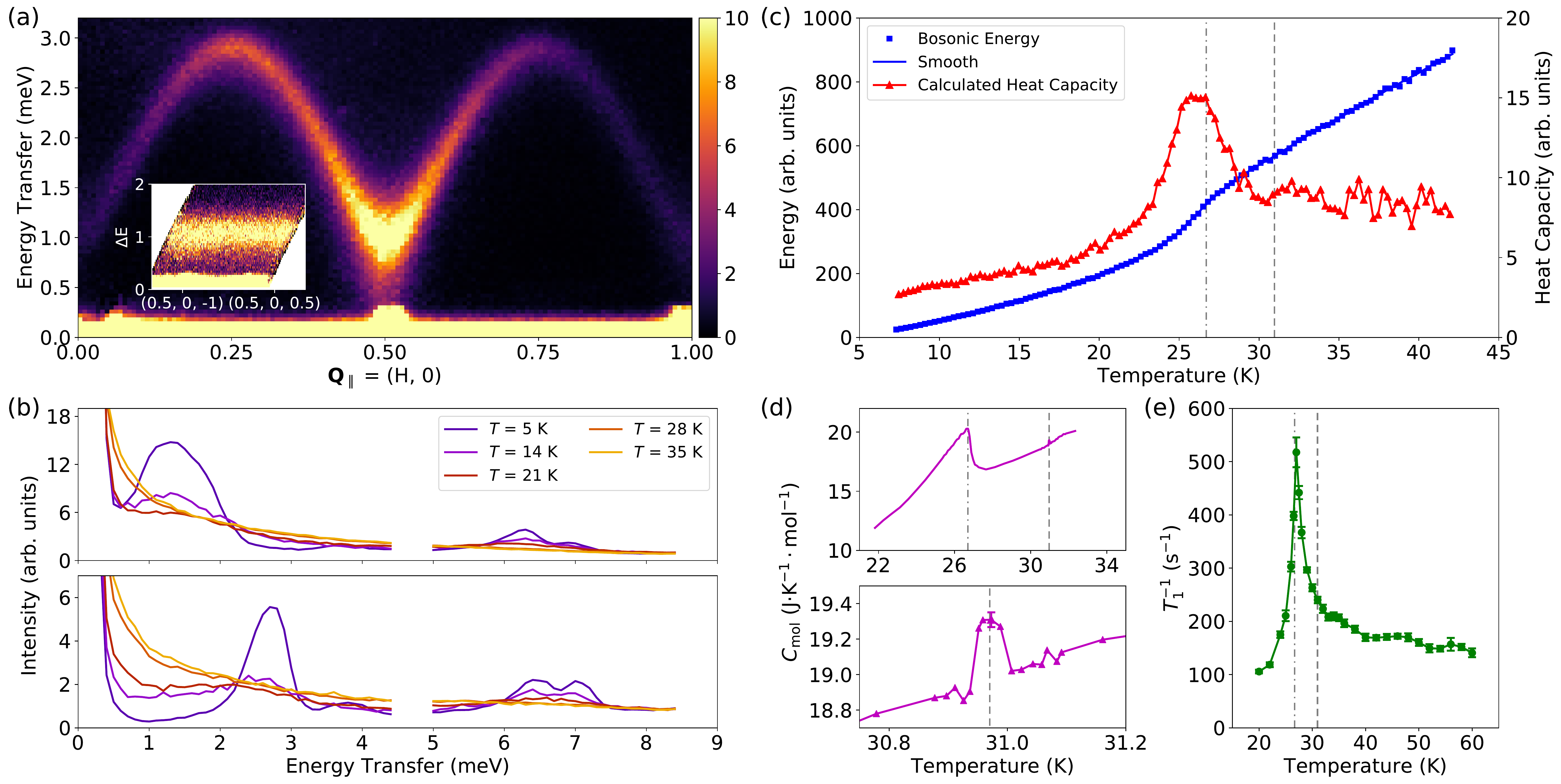}
\caption{\label{fig:3} (a) INS measurement of low-energy magnons (intensities plotted in false colors). Data at different $L$ have been combined, given the lack of dispersion along $L$ (inset). (b) Energy distribution of INS intensities at different $T$, measured at $\mathbf{q}_{\parallel}=$ (0.5, 0) (upper panel) and (0.75, 0) (lower panel). Data below and above 5 meV were obtained with incident neutron energies of 6.1 and 10.0~meV, respectively. (c) Total energy carried by magnons (estimated by weight-integrating INS intensities over [0.5, 4.5] meV), and its $T$-derivative as an estimate of the associated heat capacity, see \cite{SM} for detail. (d) High-resolution heat capacity measurement on a single crystal. Similar results have been obtained in separate runs and for two different crystals. The lower panel zooms into the part near 31 K. (e) $^{23}$Na NMR spin-lattice relaxation rate as a function of $T$, measured with $\mu_0 H = 0.75$ T $\parallel \mathbf{a^*}$. In (c-e), the dashed and dash-dotted lines indicate $T = 30.97$~K and 26.7~K, respectively.}
\end{figure*}

Figure~\ref{fig:1}(b) displays an overview of our neutron diffraction \cite{SM} result in momentum ($\mathbf{q}$) space. Long-range magnetic correlations manifest themselves as sharp peaks in the 2D ($H$, $K$) plane, at the $M$-points of the structural BZ. In the view of zigzag order \cite{Lefrancois16, Bera17}, the six peaks come from domains in which the orientations of zigzag chains differ by 120$^\circ$, as each domain contributes signal at a pair of opposite $\mathbf{q}$. Later we will show that a more suitable view of the data is that the order is of triple-$\mathbf{q}$ nature, producing diffraction signals at all the $M$-points. Turning our attention to the 3D structure of the diffraction, it is seen that the signal has a nearly $L$-independent part (``rods'' of scattering running along $L$), plus extra peaks at integer $L$. This result is presented in detail in Fig.~\ref{fig:2}(a-b), which also shows the temperature ($T$) dependence. While the integer-$L$ signal (at $H=0.5$) is much greater in amplitude at $T=5$~K, the total intensity of the rod-like signal is weaker but within an order of magnitude comparable to it. Thus there is a coexistence of 2D and 3D magnetic orders at $T=5$~K. As $T$ increases, the 3D integer-$L$ signal disappears above 26.7 K, yet the 2D rod-like signal clearly survives up to higher $T$. At 28~K we only see the 2D order [Fig.~\ref{fig:2}(a)]. We consider the (lack of) inter-layer correlation in the 2D order not merely limited by stacking faults \cite{Johnson15} manifested by the diffractions at $H=1$, since the latter are actually more structured in $L$ and with nearly no dependence on $T$.

The 2D signal [at $\mathbf{q_\mathrm{b}}$ in Fig.~\ref{fig:2}(b)] has rich structures in its $T$ dependence: (1) a local maximum at 26.7 K, presumably due to critical correlations associated with the 3D transition; (2) a slight decrease below 18~K, possibly due to subtle changes in the magnetic structure \cite{ViciuJSSC2007,Yao20,XiaoCGD2019}, which we will revisit later; (3) a smeared transition around 30~K. The smearing could be due to the limited $\mathbf{q}$ and energy resolution of neutron scattering on our large array of crystals. Figure~\ref{fig:2}(c-d) displays our resonant X-ray diffraction (RXD) \cite{SM} results from a small crystal, which beautifully complements the neutron diffraction result. An intense signal, with resolution-limited width in $H$ and perfect rod-like shape along $L$, sharply sets in below nominal $T=30$ K (the sample's actual temperature may be slightly higher, due to limited thermal shielding). Critical scattering is observed as a peak at the same $T$ near the 2D ordering $\mathbf{q}$, which unequivocally shows that the transition is of 2nd-order nature and marks spontaneous symmetry breaking. This result is significant, because if a 2D system undergoes spontaneous symmetry breaking at finite temperature, the broken symmetry must be discrete according to the Mermin-Wagner theorem. For our magnetic system, it strongly suggests that the underlying model is not Heisenberg-like or anisotropic only with a global easy-plane \cite{XiaoCGD2019}, and would be consistent with the prominence of Kitaev interactions in driving the transition.

We now turn to presenting a few aspects about the excitation spectra, which shed light on the nature of the orders and represent another main finding of this work. Analysis of the full excitation spectra will be reported elsewhere. We begin by showing low-energy magnons measured by inelastic neutron scattering (INS) \cite{SM}, deep in the 3D ordered state at $T=5$~K, along an in-plane momentum cut connecting two neighboring structural BZ centers [Fig.~\ref{fig:3}(a), the lack of dispersion along $L$ (inset) shows negligible inter-layer coupling]. As illustrated in Fig.~\ref{fig:1}(b), the cut passes through an $M$-point at (0.5,~0). Had the magnetic order been zigzag, this $M$-point would be a magnetic BZ center for one of the domains [``blue'' in Fig.~\ref{fig:1}(b)]; meanwhile, it would be a BZ \textit{corner} for the other two domains. Since (0,~0) and (1,~0) are always BZ centers regardless of domain, and because our coaligned sample must have (statistically) equal domain population, the chosen cut should manifest \textit{two inequivalent} magnon dispersions superposed on each other, but the data in Fig.~\ref{fig:3}(a) clearly disprove this. The data instead show a single magnon branch without any sign of domain superposition. Given the previous neutron diffraction evidence for zigzag order \cite{Lefrancois16, Bera17}, we interpret our result as strong evidence for a triple-$\mathbf{q}$ order, the BZ of which [Fig.~\ref{fig:1}(b)] would be fully consistent with Fig.~\ref{fig:3}(a). Indeed, a triple-$\mathbf{q}$ order can be understood as a vector sum of three 120$^\circ$-different zigzag orders (Fig.~S3 in \cite{SM}); because each zigzag order, when existing alone, would produce diffraction signals at a different $M$-point, diffraction from the triple-$\mathbf{q}$ order would be \textit{in principle} indistinguishable from the zigzag order with equal domain population. Additional indirect evidence against interpreting the order as zigzag can be found in \cite{SM}.

In recent studies \cite{Songvilay20,Lin20,Kim20} of \ch{Na_2Co_2TeO_6} powder, INS spectra of magnons have been simulated based on an assumed zigzag ground state. In light of our results, conclusions based on those analyses might need reevaluation. Our INS spectra are consistent with those of \cite{Songvilay20,Lin20,Kim20}: we see a gap below $\sim1$ meV, a lowest magnon branch up to $\sim3$ meV, and additional branches at higher energies. Figure~\ref{fig:3}(b) presents spectra at selected $\mathbf{q}$ and their $T$ dependence. We find that the high-energy branches start to form below 26.7~K, as a pile-up of intensities above 6 meV is seen at $T = 21$ K (but not at 28 K), meanwhile the low-energy intensities are depleted. Using $\mathbf{q}$-integrated INS intensities as a measure of the thermally populated boson number, we can estimate the magnons' total energy, the $T$-derivative of which provides an estimate of their heat capacity [Fig.~\ref{fig:3}(c)]. The calculated heat capacity shows a prominent peak at 26.7 K, consistent with direct measurement results [Fig.~\ref{fig:3}(d)].

It therefore seems that the order below 26.7 K satisfies a considerable portion of the leading interactions, whereas the 2D order does not. This result, in conjunction with the lack of inter-layer coupling [Fig.~\ref{fig:3}(a) inset], indicates that the 2D order is \textit{not} simply the 3D order without inter-layer correlation, and must be something fundamentally different. This understanding is consistent with our $^{23}$Na nuclear magnetic resonant (NMR) experiment \cite{SM}, which shows a pronounced peak in the spin-lattice relaxation rate $1/T_1$  at 26.7~K [Fig.~\ref{fig:3}(e)], but no pronounced anomaly upon the formation of the 2D order. The lack of 2D order's clear signature in  $1/T_1$ might be because the order is not strictly static, such that the slow dynamics wipes out its NMR signal (Fig.~S4 in \cite{SM}). We emphasize that the 2D ordering is still a genuine phase transition, as supported both by the RXD data in Fig.~\ref{fig:2}(d) and by our high-resolution heat capacity measurement \cite{SM}, which does show a tiny yet well-defined peak at $T=30.97$~K [zoom-in panel of Fig.~\ref{fig:3}(d)].

\begin{figure}[!ht]
\includegraphics[width=0.5\textwidth]{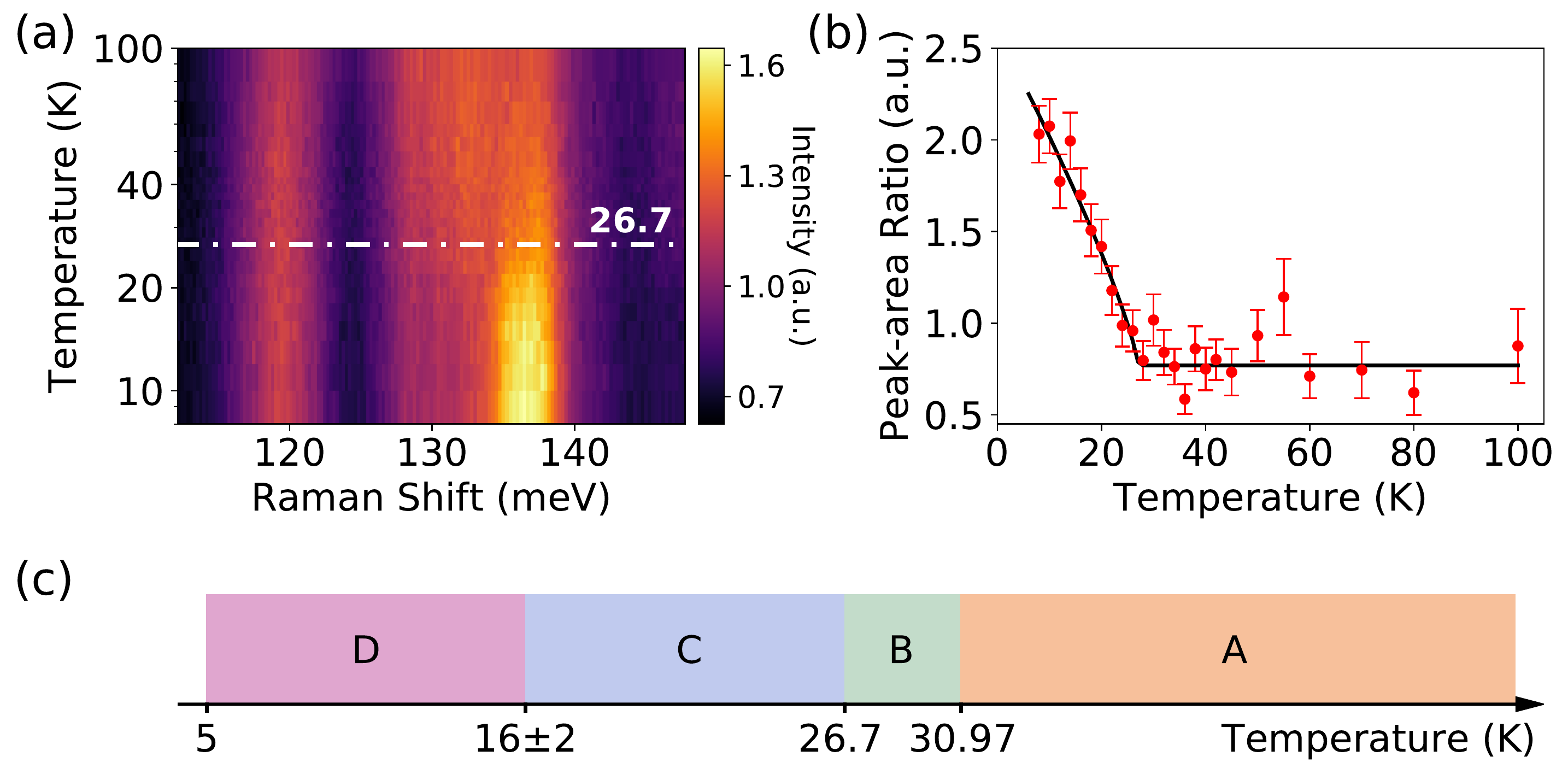}
\caption{\label{fig:4} (a) High-energy Raman spectra at different $T$ ($T$ in log scale), normalized to the peak at 119 meV. (b) $T$ dependence of area ratio between the two peaks at 136 and 130 meV (data and fits are presented in Fig.~S7 of \cite{SM}). (c) Summary of the thermal phases. A: paramagnetic; B: 2D order, no clear magnon branches; C: 2D and 3D orders, gapless excitations and magnons; D: 2D and 3D orders, gapped magnons.}
\end{figure}

The 3D order's impact goes well beyond the magnons. At much higher energies, probed by Raman spectroscopy \cite{SM} above 100 meV (Fig.~\ref{fig:4}), we still find a pronounced spectral change related to the order's formation. The high energies cannot be attributed to the pseudospin-1/2 magnetism, and although they could be related to multi-phonon scattering, we do not find any magnetoelastic effects across the 3D transition in the single-phonon spectra (Fig.~S5 in \cite{SM}). We therefore attribute the features to spin-orbit excitons \cite{SartePRB2019} and/or multi-phonon scattering assisted by electronic excitations. In either case, the excitations likely involve the orbital degrees of freedom, which further signifies the role of spin-orbit physics in driving the 3D order. In the opposite energy limit, in contrast, the 3D order is unable to immediately open up the 1 meV gap at the $M$-point [Fig.~\ref{fig:3}(b)], as the first clear indication of such a gap is not seen until $T=14$~K (detailed $T$ dependence is presented in Fig.~S6 of \cite{SM}). The persistence of gapless excitations, down to the subtle transition temperature 15-18 K shown in Fig.~\ref{fig:2}(b) and in Refs. \cite{ViciuJSSC2007,Yao20,XiaoCGD2019,Lin20}, suggests that the 3D order manifests substantial frustration upon its initial formation, until the system is cooled much more deeply to freeze out the nearly degenerate configurations. Our view of the thermal phases is summarized in Fig.~\ref{fig:4}(c).

On a final note, we discuss possible implications of our discovery of the 3D triple-$\mathbf{q}$ order in conjunction with the other phases. Similar to other multiple-$\mathbf{q}$ magnetic orders \cite{ChristensenPRB2015,GastiasoroPRB2015}, we expect the 3D triple-$\mathbf{q}$ order here to give rise to a charge (orbital) order (Fig.~S3 in \cite{SM}) that might exist on its own as a ``vestigial'' order \cite{FernandesPRB2016}. An intriguing possibility is that the 2D order is of this type -- the spin-orbit entangled pseudospin-1/2 degree of freedom might further enrich such orders by intermixing the magnetic and charge order parameters. Moreover, the triple-$\mathbf{q}$ order, formed by the superposition of three zigzag components, possesses non-zero pseudospin vorticity (Fig.~S3 in \cite{SM}) that is shared by previously proposed ``vortex'' \cite{ChaloupkaPRB2015} and ``cubic'' \cite{RousochatzakisPRL2017} orders, albeit the latter orders are characterized by the $K$- rather than the $M$-points of the BZ. We are not aware of previous reports of the triple-$\mathbf{q}$ order that we propose. The vorticity might leave distinct signatures in experiments sensitive to time-reversal symmetry breaking, \textit{e.g.}, optical Kerr rotation. In fact, the emergent vorticity may couple to external magnetic fields in a fashion similar to ferro- and/or ferrimagnetism, which has been observed \cite{Yao20}. The self-organization of vorticity might be related to the transition near 15-18 K, hence explaining the freezing of the gapless excitations below this temperature.

To summarize, our work calls for a revision of the magnetic ground state of \ch{Na_2Co_2TeO_6}, highlighting the importance of triple-$\mathbf{q}$ rather than zigzag order for navigating our determination of the microscopic model. We present systematic evidence that spin-orbit physics, highly anisotropic interactions, and magnetic frustrations are prominent characteristics of the model, which are in line with proposals related to realizing the Kitaev model. The system clearly manifests intriguingly rich physics.

\begin{acknowledgments}
We are grateful for discussions with Gang Chen, Rafael Fernandes, Giniyat Khaliullin, Haijun Liao, Yingying Peng, Natalia B. Perkins, Yiming Qiu, Kaiwei Sun, Fa Wang, Liusuo Wu, Tao Xiang, and Yang Zhao. We thank Weiliang Yao for his assistance with sample preparation and the neutron scattering experiment, which was performed at the MLF, J-PARC, Japan, under a user programme (proposal No. 2019B0062). Part of the research described in this work was performed at the Canadian Light Source, a National Research Facility of University of Saskatchewan, which is supported by the Canada Foundation for Innovation (CFI), the Natural Sciences and Engineering Research Council (NSERC), the National Research Council (NRC), the Canadian Institutes of Health Research (CIHR), the Government of Saskatchewan, and the University of Saskatchewan. Y.L. acknowledges support by the NSF of China (Grant No. 11888101) and the National Basic Research Program of China (Grant No. 2018YFA0305602). X.L. acknowledges support by the NSF of China (Grant No. 11921005) and the Strategic Priority Research Program of Chinese Academy of Sciences (Grant No. XDB28000000). W.Y. acknowledges support by the Ministry of Science and Technology of China (Grant No. 2016YFA0300504) and the NSF of China (Grant No.~51872328).
\end{acknowledgments}

\bibliography{NCTO_phase_transition}

\pagebreak
\pagebreak
\begin{center}
\textbf{\large Supplemental Materials for ``Spin-orbit phase behaviors of \ch{Na_2Co_2TeO_6} at low temperatures''}
\end{center}
\setcounter{equation}{0}
\setcounter{figure}{0}
\setcounter{table}{0}
\setcounter{page}{1}
\makeatletter
\renewcommand{\theequation}{S\arabic{equation}}
\renewcommand{\thefigure}{S\arabic{figure}}
\renewcommand{\bibnumfmt}[1]{[S#1]}


\section{Additional Data and Discussion}
\begin{figure*}[!ht]
\centering
\includegraphics[width=0.7\textwidth]{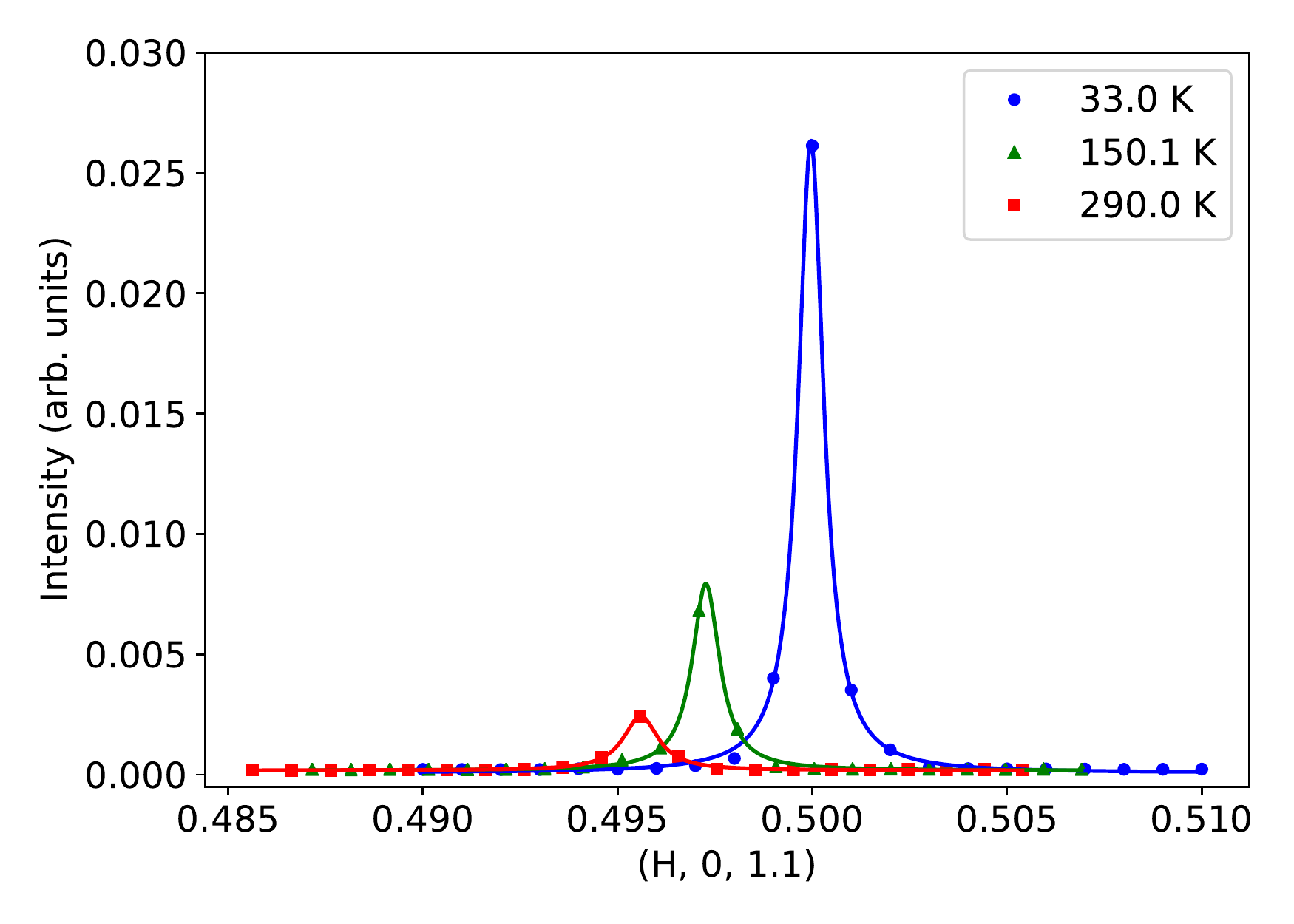}
\caption{RXD diffraction due to a superstructural modulation characterized by $\mathbf{q} = (0.5, 0, 1.1)$. The diffraction signal, resonantly enhanced at 1075.4 eV (Na $K$-edge), persists to near room temperature.}
\label{fig:S-LR_Hscan_T-dep}
\end{figure*}

\begin{figure*}[!ht]
\centering
\includegraphics[width=0.6\textwidth]{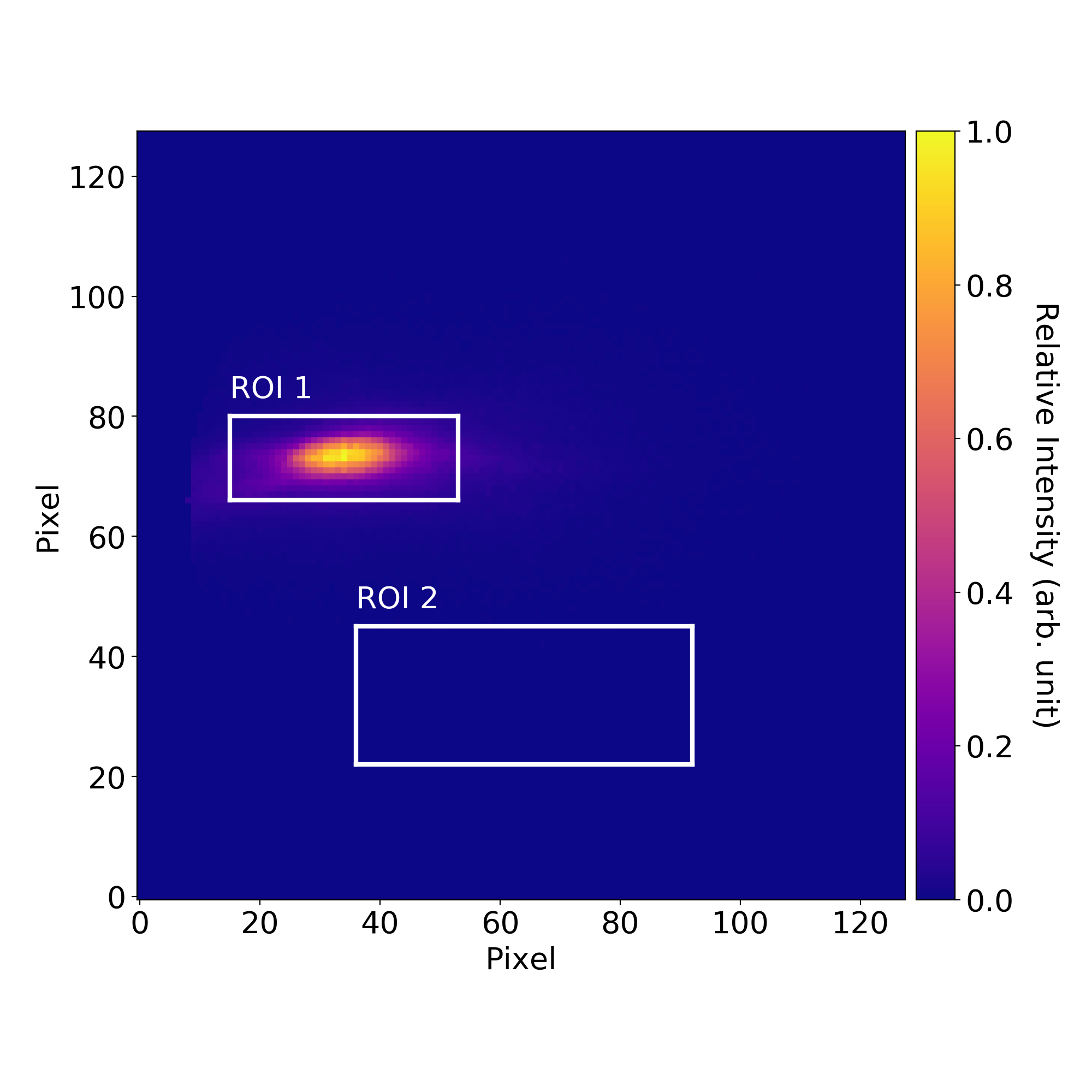}
\caption{Region of interest (ROI) on the MCP detector, in order to extract the magnetic diffraction signal at $\mathbf{q_{m(2D)}} = (0.5, 0, 0.62)$ (ROI 1) and critical scattering associated with the phase transition (ROI 2).}
\label{fig:MCP_image_ROI_demo}
\end{figure*}

\begin{figure*}[!ht]
\centering
    \includegraphics[width=0.7\textwidth]{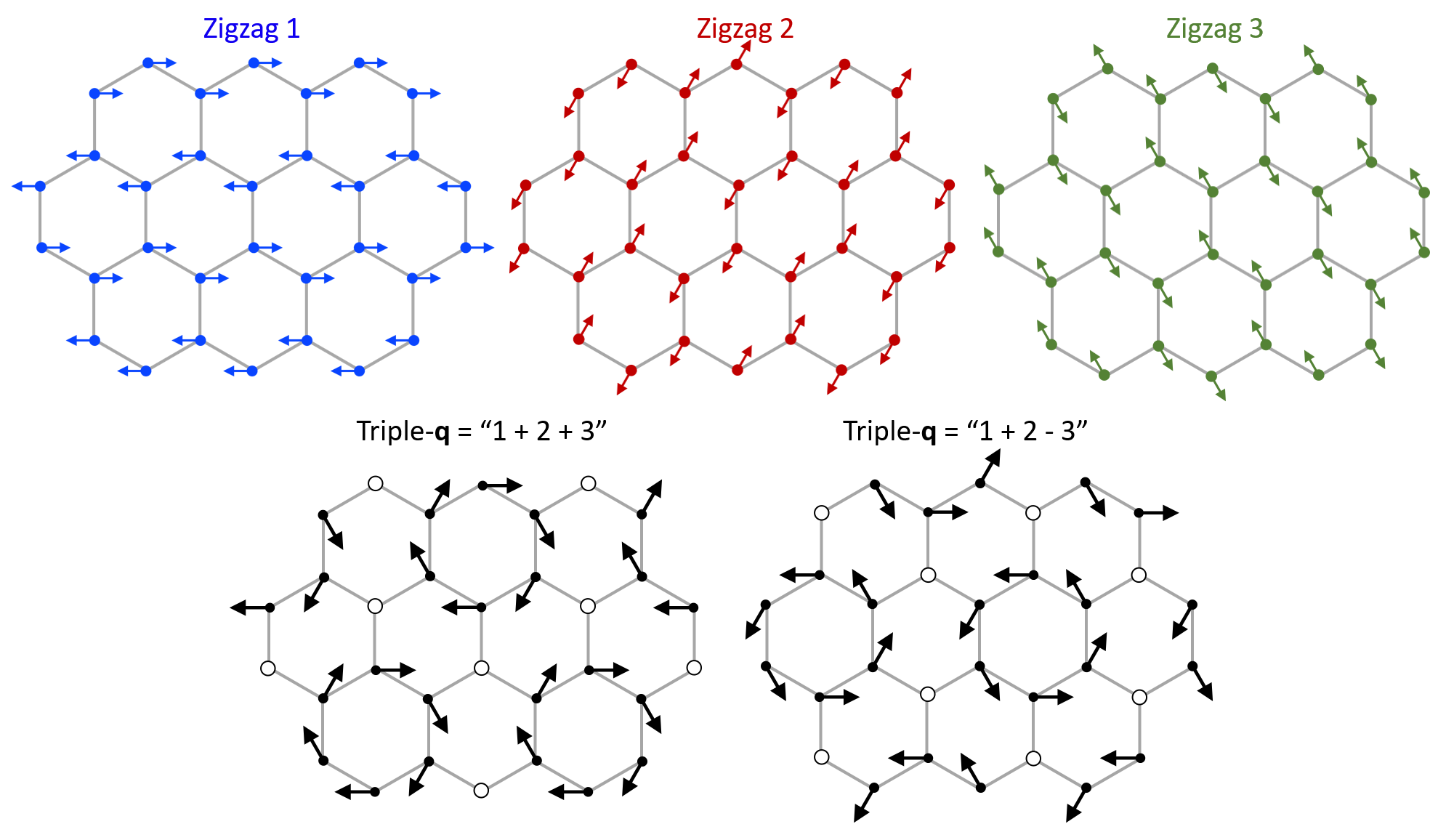}
    \caption{\label{fig:triple} Illustration of the triple-$\mathbf{q}$ order, formed by superposing three zigzag order parameters with spins lying in the $ab$-plane. Empty circles indicate spin-less sites. A total of eight domains can be expected in the triple-$\mathbf{q}$ phase, all preserving the $C_3$ symmetry, whereas the ``vorticity'' of the spin texture can be reversed by altering the sign of one of the three constituent zigzag components. When the zigzag patterns have antiferromagnetic out-of-plane spin canting, the spin-less sites will have $c$-axis moments that also form an antiferromagnetic order (N\'{e}el-type) on a honeycomb lattice that is twice of the original lattice.}
\end{figure*}

Figures \ref{fig:S-LR_Hscan_T-dep} - \ref{fig:Raman_fitting} display additional data mentioned in the main text. Description of the measurements can be found in the captions.

In additional to the key evidence against the zigzag interpretation of the 3D magnetic order, based on Brillouin zones (BZs) of magnetic excitations as presented in the main text, we have found three additional pieces of indirect evidence:

(1) With the understanding that the zigzag order features staggered magnetic moments pointing along zigzag chains \cite{Lefrancois16, Bera17}, a sufficiently strong in-plane magnetic field is expected to lead to unequal zigzag-domain population when a crystal is field-cooled into the ordered state, and such unequal domain population ought to affect the magnetic response of the sample. However, as mentioned already in \cite{Yao20}, field-cooling crystals with magnetic fields up to 7 Tesla parallel to either the $\mathbf{a}$ or the $\mathbf{a}^*$ direction, then turning off the field and measuring the sample's variable-field magnetization, does not result differently from the corresponding measurements after preparing the sample's low-temperature state with zero-field-cooling.

(2) In the presence of magnetoelastic coupling, a zigzag order is expected to lead to $C_3$ symmetry breaking of the crystal lattice, resulting in splitting of diffraction peaks with high in-plane indices in the ordered state. Within the experimental resolution, our single-crystal neutron diffraction data do not show any indication of such splitting (Fig.~\ref{fig:ND_maps_cuts}).

(3) In a zigzag-ordered state, magnetoelastic coupling is also expected to lift the energy degeneracy of zone-center optical phonons that belong to the $E_1$ and $E_2$ irreducible representations of the $D_6$ point group (or P$6_322$ space group) of \ch{Na_2Co_2TeO_6}. However, our high-resolution polarized Raman spectroscopy measurements of single crystals of \ch{Na_2Co_2TeO_6} show no phonon splitting at all at $T=10$~K compared to 40~K (Fig.~\ref{fig:Raman_more}). The spectra are simply identical.

\section{Experimental Methods}
\subsection{Resonant X-Ray Diffraction (RXD)}
Our resonant x-ray diffraction (RXD) experiment was performed at the soft-x-ray beamline 10ID-2 at Canadian Light Source (CLS). A high-quality single crystal of \ch{Na_2Co_2TeO_6} of about 3 mm $\times$ 3 mm lateral dimensions was used for the experiment (sample photo shown in Fig.~\ref{fig:X-ray_sample_photo}).  Incident x-ray photons of energy 779.5 eV (cobalt $L_3$-edge) was used to maximize the diffraction signal at low temperature at $H=0.5$ (Figs. \ref{fig:XAS} and \ref{fig:H05_E-dep_L3}), the magnetic nature of which has been verified by incident-photon polarization analysis [Fig.~\ref{fig:pol-dep_LTHT}]. A microchannel plate (MCP) detector was used in the experiment, which had a photon acceptance diameter of 2.5 cm. The distance from the sample to the MCP detector is 300 mm. Data presented in Fig.~2(c) of the main text were obtained after reconstructing a sequence of MCP images, taken in slightly different scattering geometries, into a volume data and taking the $K=0$ slice of the volume (Fig. \ref{fig:M-LR_H0L-plane}). All key results have been confirmed in a separate experiment on a different crystal.

\subsection{Neutron Diffraction and Inelastic Scattering}
Our neutron scattering experiment was performed on the 4SEASONS time-of-flight spectrometer at the MLF, J-PARC, Japan \cite{Kajimoto11}. The spectrometer’s multiple-$E_\mathrm{i}$ capability \cite{Nakamura09} enabled us to simultaneously obtain data in different energy ranges with different energy resolutions. All data were acquired with chopper frequency of 150 Hz. In this condition and given our sample's diameter of about 25 mm, the calculated energy resolutions at the elastic condition are about 0.19, 0.31, and 0.58 meV (full width at half maximum) for incident energy $E_\mathrm{i} = 4.1$, 6.1, and 10.0 meV, respectively. Neutron diffraction and inelastic neutron scattering (INS) results were obtained in the same run, by selecting the energy transfers ($\Delta E$, zero for diffraction) based on the measured scattering time of flight.

About a total of 2 grams of high-quality single crystals of \ch{Na_2Co_2TeO_6} were used for the experiment. The crystals were co-aligned on aluminum plates with a hydrogen-free adhesive, such that ($H$,~0,~$L$) are in the horizontal plane. According to a rocking curve measured on the (1,~0,~0) nuclear Bragg reflection , the mosaic spread of the entire sample is about 2.1$^\circ$ (full width at half maximum). For measurements that produced intensity maps of two or more dimensions at fixed $T$, the sample was rotated over a 75$^\circ$ range in 0.5$^\circ$ steps, and intensity maps were cut out from the acquired four-dimensional data set; for detailed $T$ dependence measurements, the sample was set to specific orientations designed for obtaining scattering signals at the desired $\mathbf{q}$ and energy. Data were reduced and analyzed with the Utsusemi \cite{Inamura13} and Horace \cite{Ewings16} software packages. When extracting the scattering intensity, a small range in $\mathbf{q}$ and energy surrounding the designated $(\mathbf{q}, E)$ is integrated over. The specific conditions for obtaining the data presented in figures of the main text are summarized in Table~\ref{tab:INS_integral_range}.

\begin{table*}
\caption{\label{tab:INS_integral_range} Detailed conditions for neutron scattering data reduction used for figures in the main text.}
\begin{ruledtabular}
\begin{tabular}{lrrrrr}
Data & $H$ range (r.l.u.) & $K$ range (r.l.u.) & $L$ range (r.l.u.) & $E_\mathrm{i}$ (meV)\footnote{Incident neutron energy.} & $\Delta E$ range (meV)\footnote{Neutron energy transfer.}\\
\hline
Fig. 1(b) & - & - & - & 4.1 & [-0.25,~0.25]\\
Fig. 2(a) & - & [-0.03,~0.03] & - & 4.1 & [-0.1,~0.1]\\
Fig. 2(b) $\mathbf{q}_\mathrm{a}$ & [0.45,~0.55] & [-0.05,~0.05] & [-0.035,~0.055] & 10.0 & [-0.19,~0.19]\\
Fig. 2(b) $\mathbf{q}_\mathrm{b}$ & [-0.54, -0.46] & [-0.04,~0.04] & [0.435,~0.685] & 4.1 & [-0.1,~0.1]\\
Fig. 3(a) & - & [-0.05, 0.05] & all & 4.1 & -\\
Fig. 3(b) upper panel below 5 meV & [0.4, 0.6] & [-0.1, 0.1] & all & 6.1 & -\\
Fig. 3(b) upper panel above 5 meV & [0.4, 0.6] & [-0.1, 0.1] & all & 10.0 & -\\
Fig. 3(b) lower panel below 5 meV & [0.65, 0.85] & [-0.1, 0.1] & all & 6.1 & -\\
Fig. 3(b) lower panel above 5 meV & [0.65, 0.85] & [-0.1, 0.1] & all & 10.0 & -\\
\end{tabular}
\end{ruledtabular}
\end{table*}

At a given temperature, the total energy carried by bosonic excitations of magnetic origin can be calculated as
\begin{equation}
\mathcal{E}(T) \propto \int \text{DOS}(\omega, T)n(\omega, T)\hbar\omega \mathrm{d} {\omega}
\end{equation}
where $\text{DOS}(\omega, T)$ is the density of states. As an approximation, we consider
\begin{equation}
\text{DOS}(\omega, T) \propto \int I(\mathbf{q}, \omega, T) \mathrm{d} {\mathbf{q}},
\end{equation}
where $I$ is the measured INS intensity, and the integration runs over the BZ. The approximation neglects difference in the INS dynamical structure factor of different bosonic modes, and of the same mode in different BZs. $n(\omega, T)$ is the Bose-Einstein distribution
\begin{equation}
    n(\omega, T) = \frac{1}{e^{\hbar \omega / k_B T} - 1}.
\end{equation}
We can write the effective bosonic energy ($\tilde{\mathcal{E}}$) as
\begin{equation}
    \tilde{\mathcal{E}}(T) = \int\left\{\int I(\mathbf{q}, \omega, T) \mathrm{d} {\mathbf{q}}\right\} n(\omega, T)\hbar\omega \mathrm{d} {\omega}.
\end{equation}

Figures \ref{fig:int_BZ_6meV_c2w_rebinned} and \ref{fig:dU_6meV} presents the above integrations, using our $E_\mathrm{i}=6.1$ meV INS data. The $\mathbf{q}$ integration range is chosen to be greater than a single BZ zone ($H \in [-0.9,~1.5],~K \in [-0.9,~0.9]$, $L$ with all available data), in order to improve data statistics. The energy integration range is from 0.5 meV to 4.5 meV, where the lower energy bound is chosen to avoid diffraction intensities. In principle, the upper bound should be chosen as high as possible, but due to the properties of Bose-Einstein distribution, high-energy excitations contribute much less to the integration than low-energy ones, and can be neglected in this calculation. The final outcome of the calculation is presented in Fig. 3(c) of the main text. To further calculate the associated heat capacity, we take the $T$ derivative of the bosonic energy. This is done by first smoothing the energy vs. $T$ using a Savitzky–Golay filter of window size 17 and polynomial order 3, and then taking derivative of the smoothed data.

\subsection{Raman Scattering}
Our Raman experiments were performed in a confocal back-scattering geometry using a Horiba Jobin Yvon LabRAM HR Evolution spectrometer equipped with 1800 lines/mm grating and a liquid-nitrogen-cooled CCD detector. During the measurements, the samples were kept in a liquid-helium flow cryostat (ARS) under an ultrahigh vacuum ($\sim 10^{-8}$ torr). The data in Fig.~4 of the main text were obtained using a He-Ne laser with $\lambda$ = 632.8 nm for excitation, with linear polarization of both incident and scattered photons parallel to the $\mathbf{a}^*$ direction.

As the primitive cell of \ch{Na_2Co_2TeO_6} contains two formula units, one expects (from factor-group analysis) at least a total of 19 Raman-active optical phonons involving the vibration of Co, Te, and O, which can be expressed as irreducible representations of the $D_6$ symmetry group: 2 $A_1$, 8 $E_1$, and 9 $E_2$ modes (plus 1 $A_1$ and 1 $E_1$ as acoustic phonons and additional $A_2$, $B_1$, and $B_2$ modes). The Na atoms are subject to disordered partial occupation of different Wyckoff sites \cite{ViciuJSSC2007}, and they further complicates the phonon spectra. Nevertheless, by performing Raman spectroscopy on a crystal's side surface (prepared after the method in \cite{Ren15}), as shown in Fig. \ref{fig:Raman_more}(a), one is able to detect the doubly-degenerate $E_1$ and $E_2$ modes. These modes are expected to undergo mode splitting in the magnetically ordered state, if the order breaks the $C_3$ rotational symmetry, as expected for the zigzag order. Against this expectation, we find the spectra to be nearly identical between $T=10$~K and 40~K. This result suggests that either the magnetoelastic coupling is too small to be observed, or the magnetic ground state preserves the $C_3$ symmetry, \textit{i.e.}, is not zigzag.

For comparison, we have also performed Raman measurements of \ch{Na_2Co_2TeO_6} and isostructural but non-magnetic \ch{Na_2Zn_2TeO_6} at room temperature, using different laser lines ($\lambda = 514.5$ and 633 nm) for excitation [Fig. \ref{fig:Raman_more}(b)]. The measurements, performed on crystal surfaces parallel to the $ab$ plane with incident and scattered photon polarization along $\mathbf{a}^*$, same as in Fig. \ref{fig:Raman_fitting} and Fig. 4(a) in the main text, shows that the high-energy features are genuine Raman scattering (\textit{i.e.}, not photo-luminescence). Importantly, the multi-phonon features of \ch{Na_2Zn_2TeO_6} has a much simpler spectral shape than the signals of \ch{Na_2Co_2TeO_6}, consistent with our understanding that the latter additionally involves electronic orbital excitations.

\subsection{High-Resolution Calorimetry}
Ac-temperature calorimetry (ac calorimetry in short) has long been used as a common method to measure heat capacity since it was first proposed by Sullivan and Seidel \cite{Sullivan68}. This method employs a heater and a thermometer attached to a sample which has weak thermal connection with a thermal bath. An ac heating excitation current is produced by the heater, while the real-time temperature oscillation of the sample is measured at the same time. At appropriate heating excitation current frequencies, the oscillation amplitude $\Delta T$ can be used to determine the heat capacity:
\begin{equation}
C = \frac{\dot{Q}_0}{2\omega\Delta T}
\label{eq:specific_heat}
\end{equation}
where $Q_0$ is the heating power amplitude, and $\omega$ is the heating power angular frequency (twice the one of heating excitation current). Schematic diagram and a photo of our actual sample device are shown in Fig.~\ref{fig:specific_heat_1}(a, b). The measurement procedure is described as follows:

\begin{enumerate}
    \item Before the fabrication of the device, heater’s resistance was measured and thermometers were calibrated.
    \item The sample was glued with GE varnish on two parallel tightened fishing lines attached to the sample holder. Spiral manganin wires were then attached to the electrodes on the device with silver paint.
    \item The sample holder was mounted on a 4 K refrigerator and cooled down to 20 K.
    \item Sinusoidal voltage of 0.07V was generated using SR-830 lock-in amplifier and then applied over the heater ($\sim 1~\text{k}\Omega$). Simultaneously, the resistance of the thermometer was measured using four-probe method with SR-830 at a frequency of 1 kHz.
    \item To determine the appropriate heating excitation current frequency, frequency scans were conducted, the results of which are displayed in Fig. \ref{fig:specific_heat_1}(e). We used a heating excitation current frequency of 0.02 Hz for our measurements. The methods for both frequency and temperature scans are elaborated in procedure step 6.
    The methods for both frequency and temperature scans are elaborated in 6.
    \item Resistance of the thermometer was recorded during time periods each lasting around 1300 seconds and then translated into temperature according to the calibration (see upper panel of Fig. \ref{fig:specific_heat_1}(d)). Out of each sampling, data sets lasting integral multiples of the heating period were selected and processed with fast Fourier transformation (FFT), see bottom panel of Fig. \ref{fig:specific_heat_1}(d). As the results of each sampling showed, the peak in the frequency spectrum was exactly the heating frequency. We can then obtain the value of heat capacity according to Eq.~(\ref{eq:specific_heat}).
    \item In-situ re-calibration of the thermometer was conducted after the temperature scans (photo of the re-calibration device is shown in Fig. \ref{fig:specific_heat_1}(c)). This re-calibration is necessary in order to ensure accurate temperature reading of the result in Fig.~4(d) of the main text.
\end{enumerate}

We have furthermore repeated the measurement on a separate cooling of the same crystal, as well as on a different crystal, showing similar results. We therefore conclude that the peak at 30.97 K, albeit small, is genuinely a physical effect.

\subsection{$^{23}$Na Nuclear Magnetic Resonance (NMR)}
Our $^{23}$Na (\emph{I} = 3/2,  $^{23}$$\gamma$ = 11.262~MHz/T) NMR spectra were obtained by frequency sweeps, with the standard spin-echo sequence.
The spin-lattice relaxation rates 1/$^{23}$\emph{T}$_1$ were measured by the inversion-recovery method, with a $\pi$/2 pulse
as the inversion pulse. The recovery curve is fitted by the stretched exponential function, $I(t)/I(0) = 1-b[e^{-(t/T_1)^\beta}+9e^{-6(t/T_1)^\beta}]$, where $\beta$ is the stretching factor.
$\beta$ is found to be 1 at temperatures above 15~K, and less than 1 when cooled below 15~K (far below $T_N$), which indicates high quality of the single crystal.

\begin{figure*}[!ht]
\centering
    \includegraphics[width=0.7\textwidth]{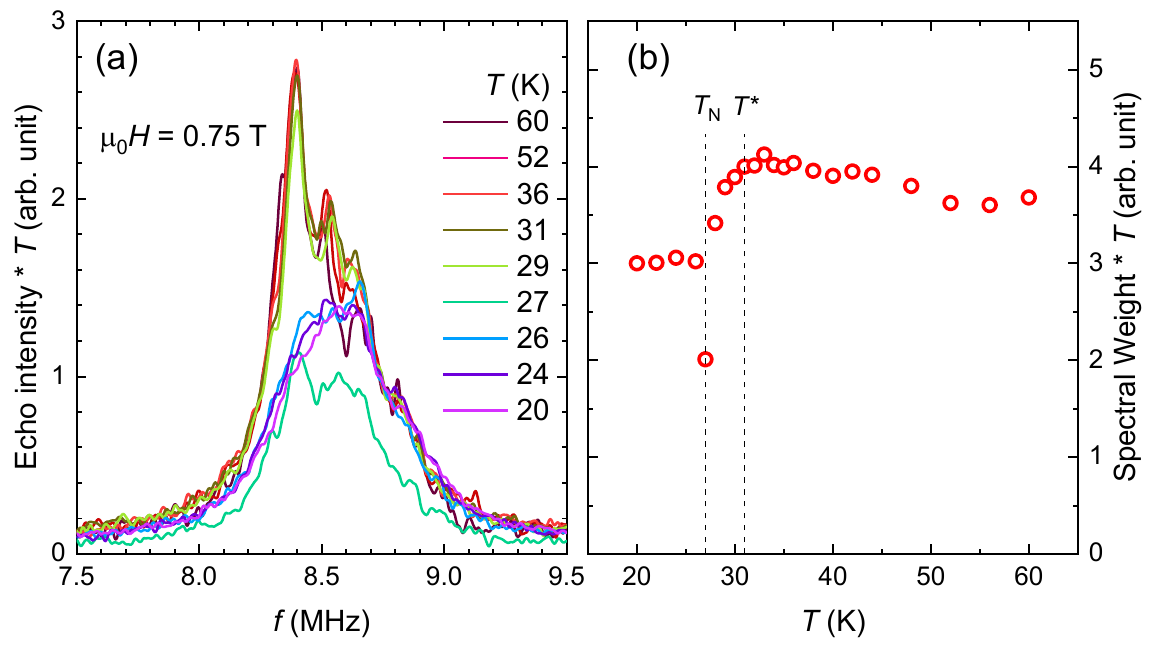}
    \caption{\label{fig:NMR_SM} (a) $^{23}$Na NMR spectra at selected temperatures, measured under a low field of 0.75~T applied along the $a^*$ direction. The Echo intensity is multiple by temperature for correction of nuclear magnetization. (b) Integrated spectral weight as a function of temperature. $T^*\approx$~31~K labels the onset temperature for the loss of the NMR spectral weight, and $T_\mathrm{N}\approx$~27~K labels the 3D magnetic transition temperature.}
\end{figure*}

\begin{figure*}[!ht]
\centering
\includegraphics[width=0.9\textwidth]{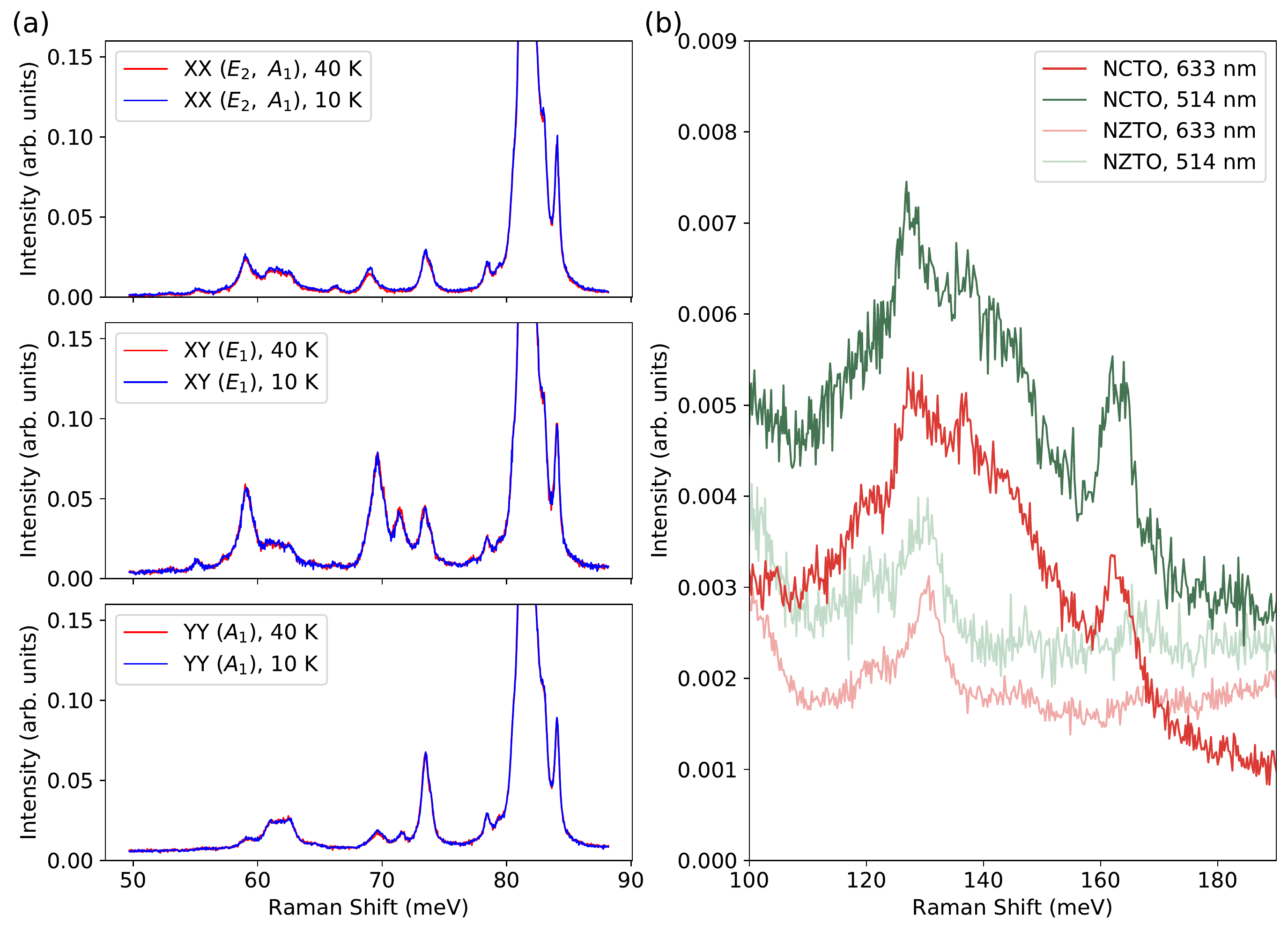}
\caption{(a) Raman spectra of \ch{Na_2Co_2TeO_6} obtained on a crystal's side surface in three polarization configurations, which detect phonon modes corresponding to different irreps of the $D_6$ group (see text in Experimental Methods). The apparent similarity between different configurations is due to polarization leakage and presence of highly disordered Na. The spectra have been normalized to the highest peak at 81 meV (out of range). (b) Raman spectra of \ch{Na_2Co_2TeO_6} and \ch{Na_2Zn_2TeO_6} measured with different lasers at room temperature.}
\label{fig:Raman_more}
\end{figure*}

\begin{figure*}[!ht]
\centering
\includegraphics[width=0.8\textwidth]{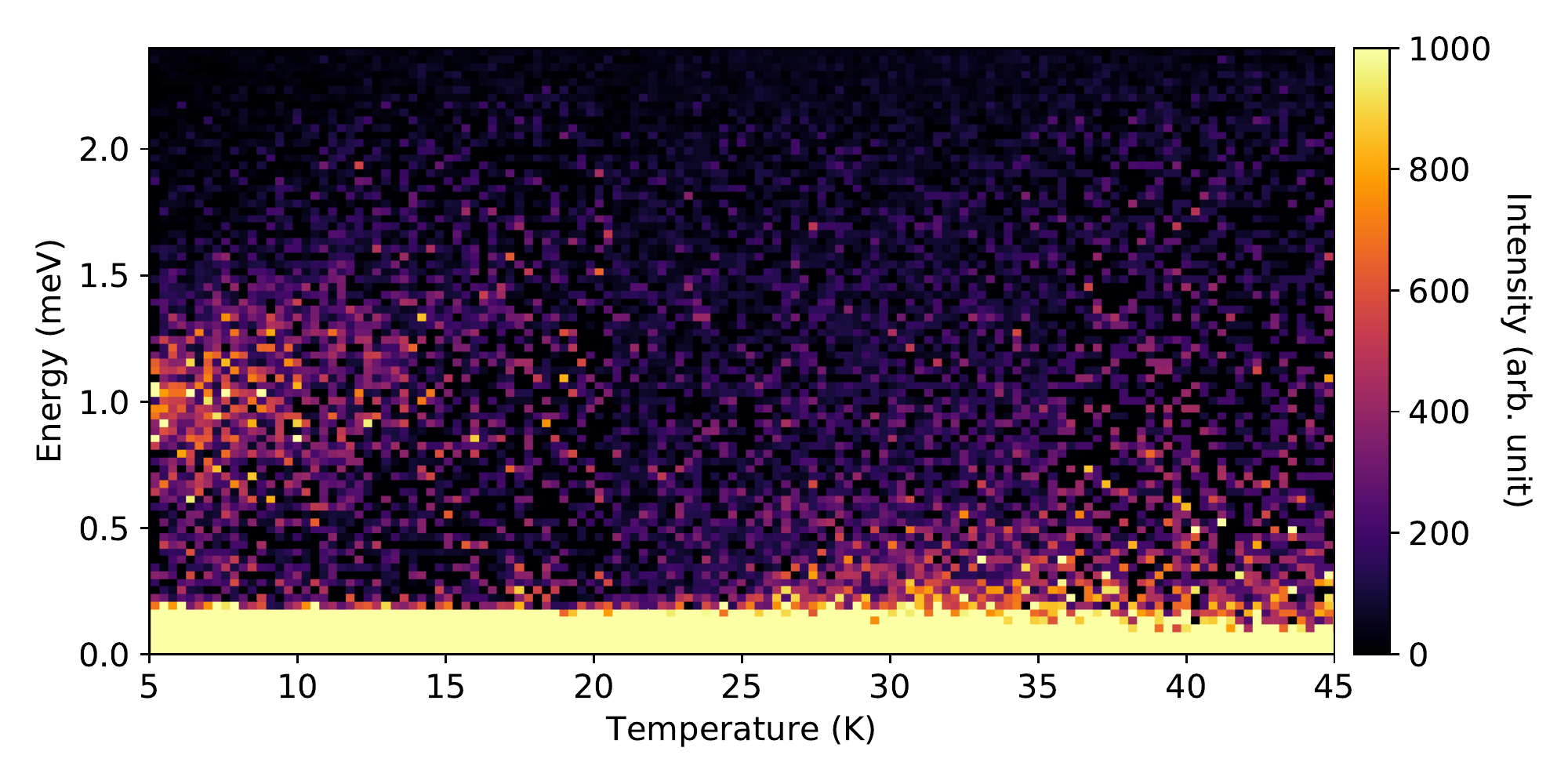}
\caption{Temperature dependence of INS spectrum acquired at the band bottom [$\mathbf{q} = (0.5, 0)$] of the lowest-energy spin-wave branch. Gapped ($\sim1$ meV) excitations are observed below about 16 K. The integral range are $H \in [0.47,~0.53],~K \in [-0.03,~0.03]$, and all $L$.}
\label{fig:gap_opening_3meV_rebinned}
\end{figure*}

\begin{figure*}[!ht]
\centering
\includegraphics[width=0.8\textwidth]{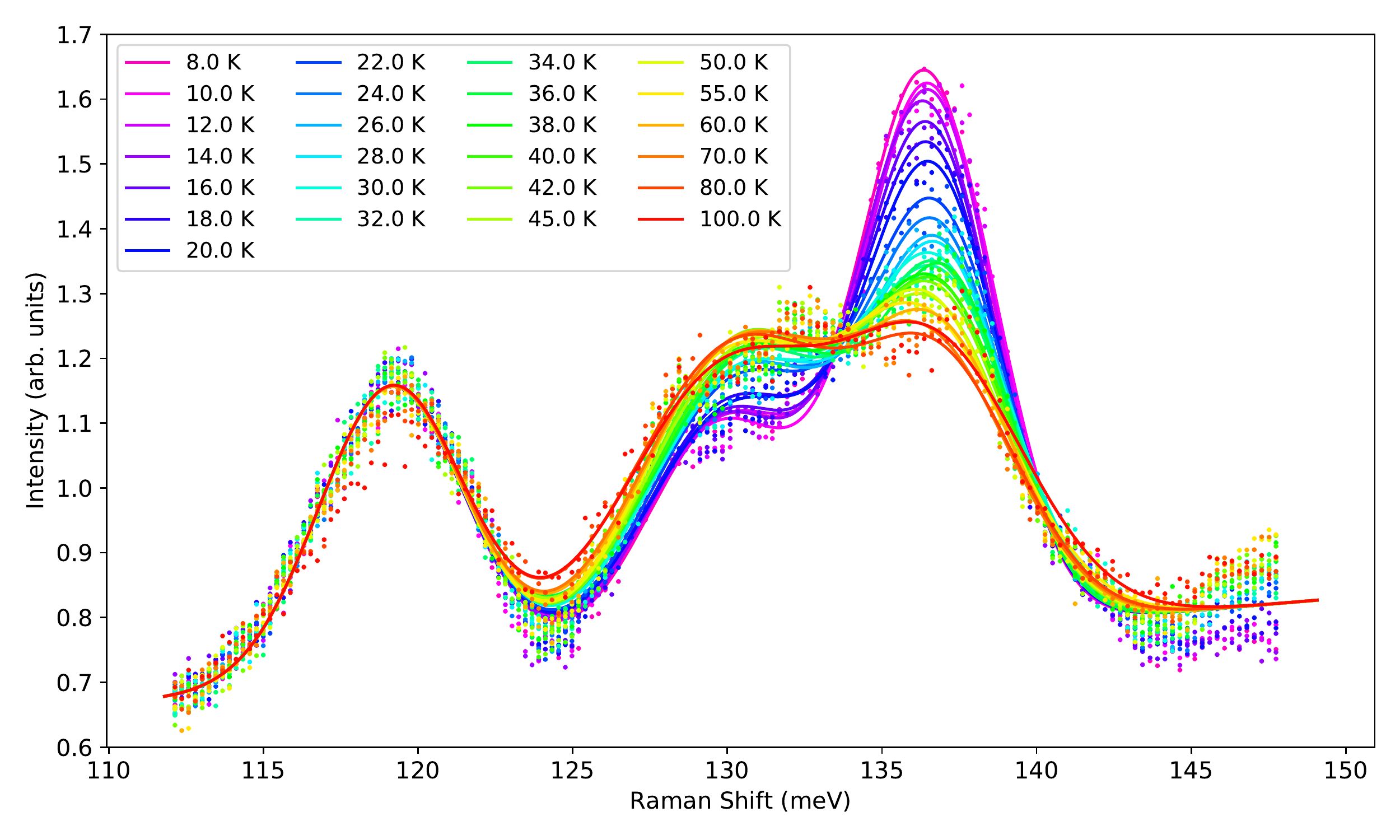}
\caption{Normalized Raman spectra (dots) and fits (lines) for the three peaks (located at 119 meV, 130 meV, and 136 meV).}
\label{fig:Raman_fitting}
\end{figure*}

\begin{figure*}[!ht]
\centering
\includegraphics[width=0.85\textwidth]{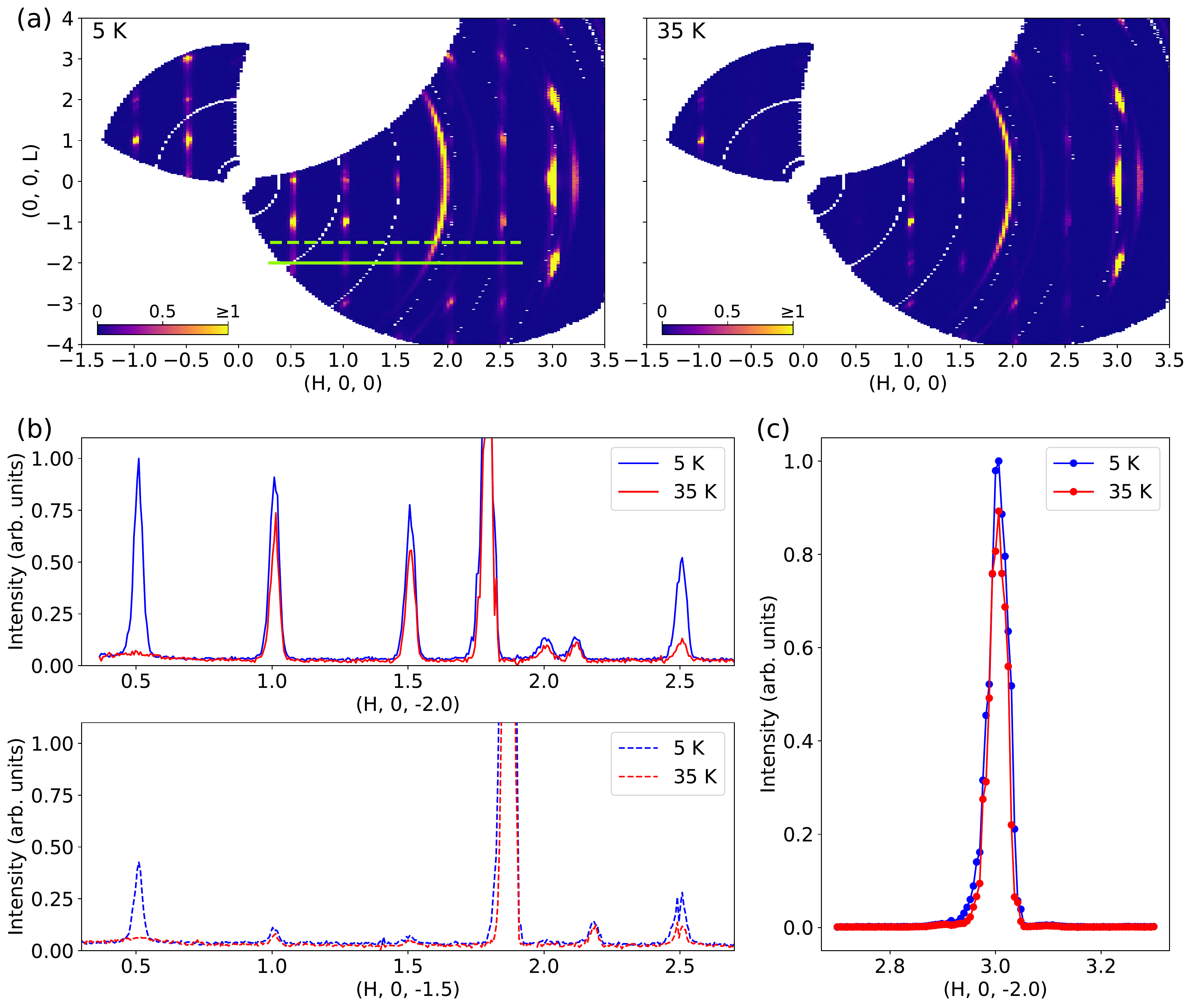}
\caption{(a) Neutron diffraction in the $(H,~0,~L)$ plane at 5 K and 35 K. Incident neutron energy is 19.4 meV. Solid line: $L = -2.0$ cut; dashed line: $L = -1.5$ cut. (b) Intensities along the two cuts at 5 K and 35 K. Integral range: $\Delta K = 0.2$, $\Delta L = 0.2$. (c) $H$-cut through the structural Bragg peak at $\mathbf{q} = (3,~0,~-2)$ at 5 K and 35 K.}
\label{fig:ND_maps_cuts}
\end{figure*}

\begin{figure*}[!ht]
\centering
\includegraphics[width=0.5\textwidth]{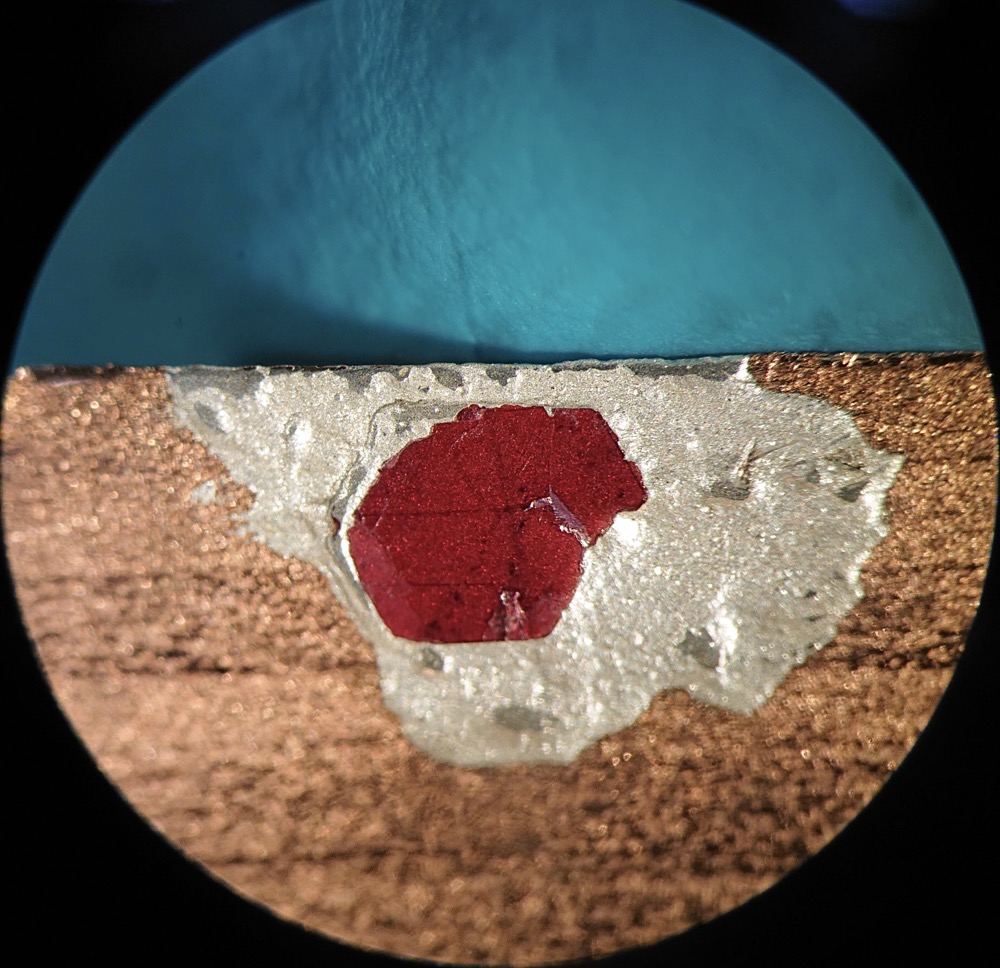}
\caption{High-quality single crystal of \ch{Na_2Co_2TeO_6} used in our RXD experiment. Lateral dimensions: $\sim$ 3 mm $\times$ 3 mm.}
\label{fig:X-ray_sample_photo}
\end{figure*}

\begin{figure*}[!ht]
\centering
\includegraphics[width=0.8\textwidth]{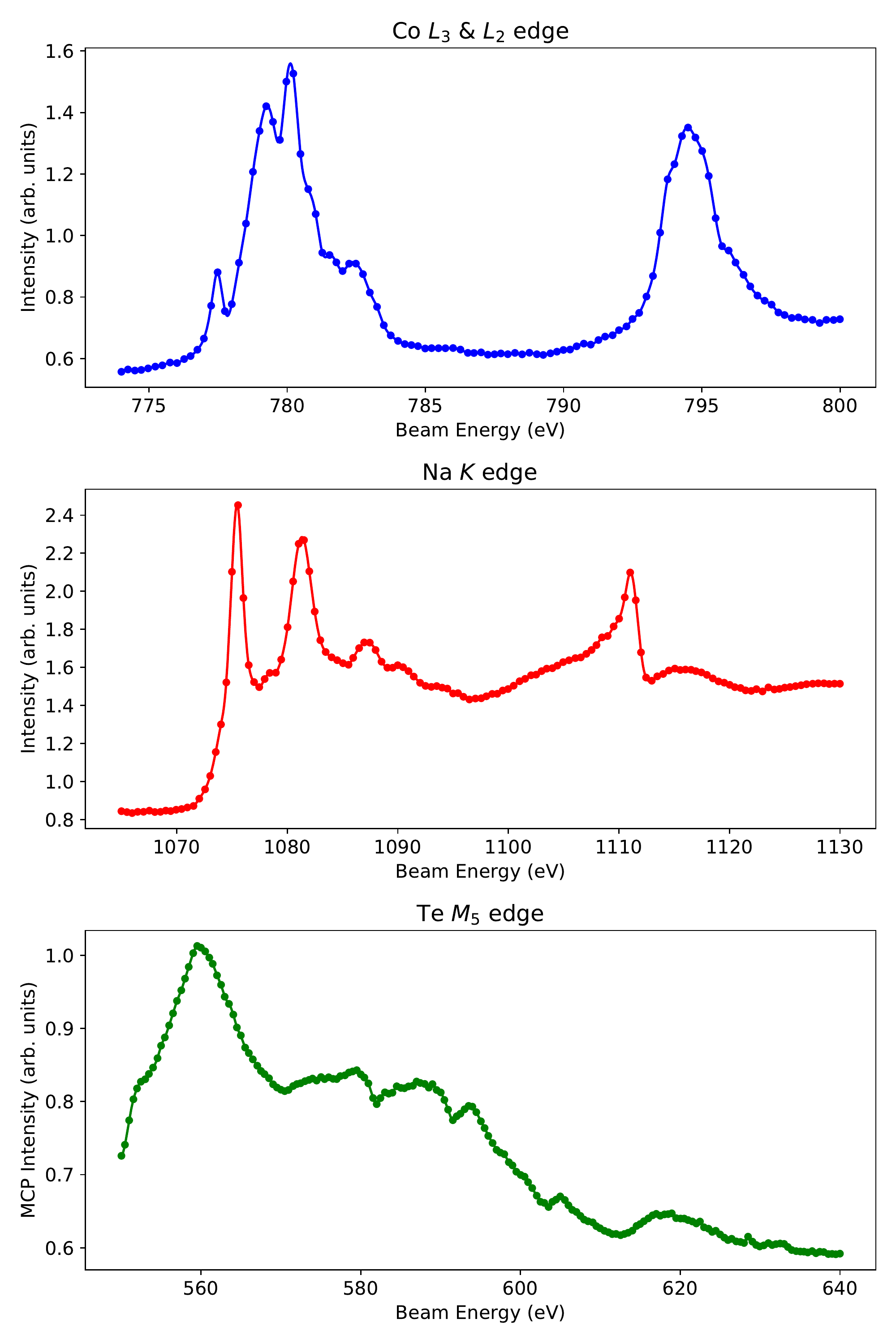}
\caption{X-ray absorption spectra of three elements in \ch{Na_2Co_2TeO_6}, measured in total fluorescence yield mode by the MCP detector upon changing the incident X-ray energy.}
\label{fig:XAS}
\end{figure*}

\begin{figure*}[!ht]
\centering
\includegraphics[width=0.8\textwidth]{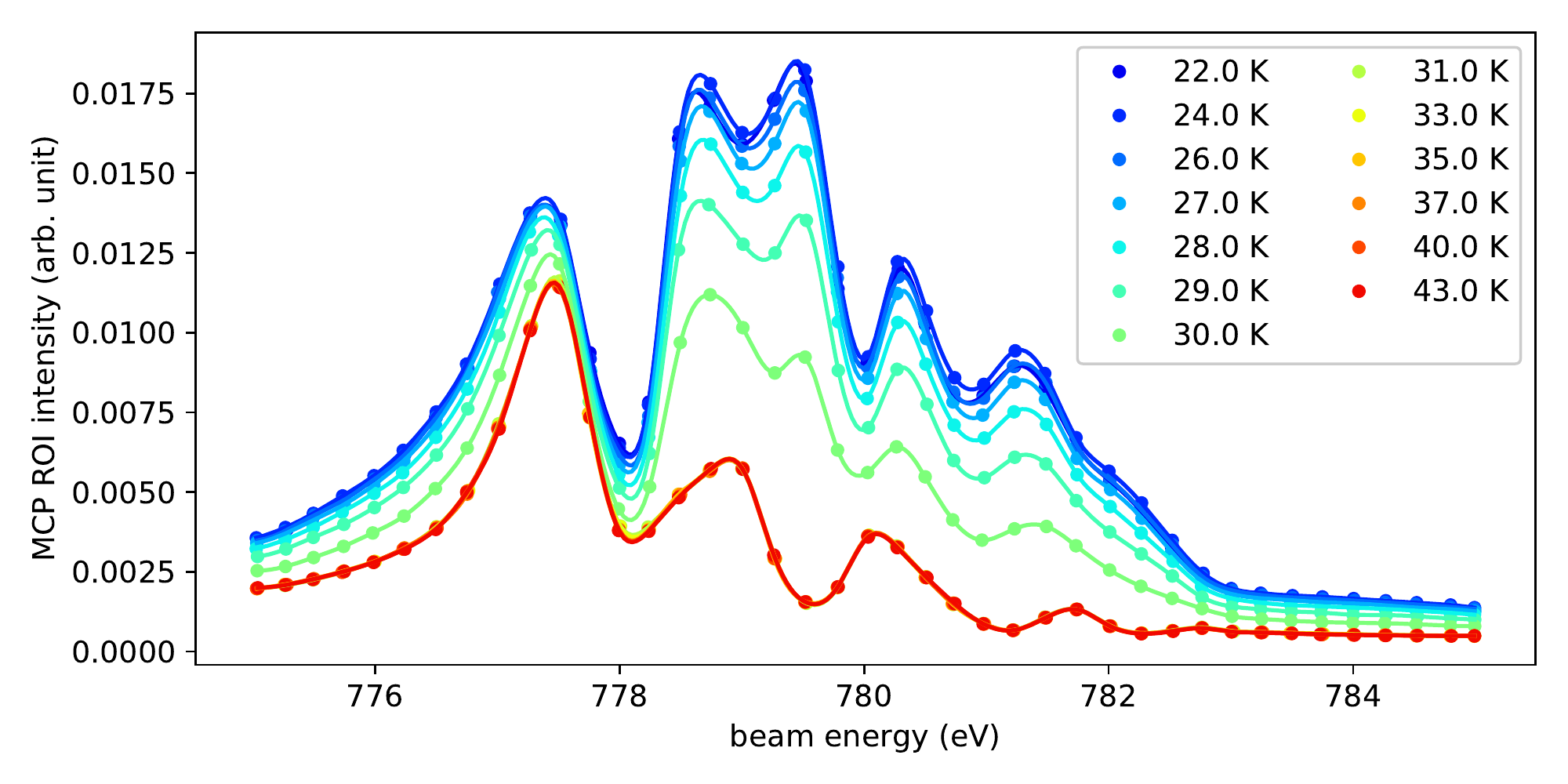}
\caption{Energy dependence of the diffraction signal at $\mathbf{q} = (0.5, 0, 0.64)$ near the cobalt $L_3$-edge. The incident polarization is horizontal ($\pi$-incident).}
\label{fig:H05_E-dep_L3}
\end{figure*}

\begin{figure*}[!ht]
\centering
\includegraphics[width=0.6\textwidth]{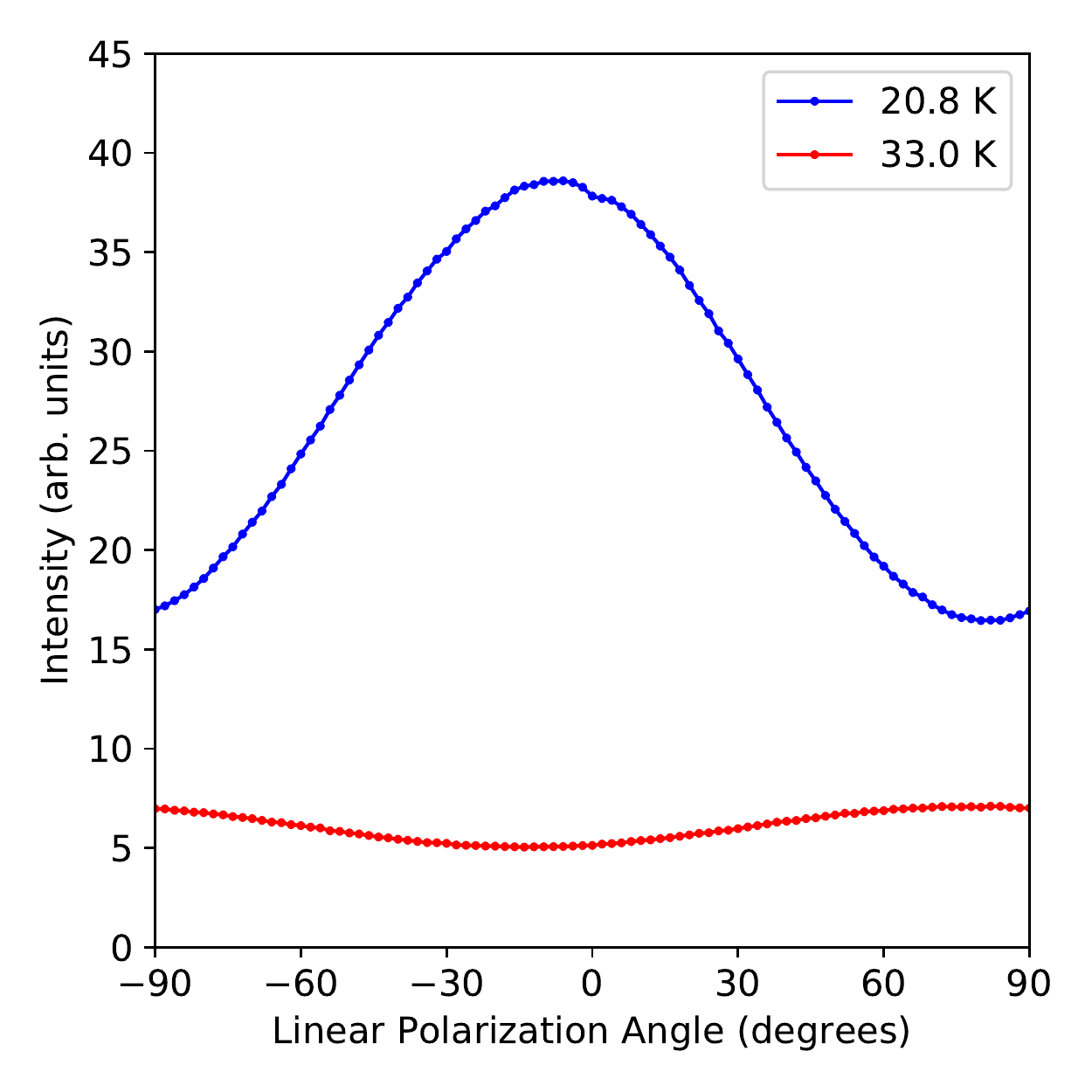}
\caption{Incident-photon polarization dependence of diffraction signals at $\mathbf{q} = (0.5, 0, 0.64)$. The linear polarization angle $0^\circ$ corresponds to horizontal polarization ($\pi$-incident), while $\pm 90^\circ$ corresponds to vertical polarization ($\sigma$-incident). While the high-temperature signal is favored by the $\sigma$-incident condition, consistent with its charge origin (superstructure in the lattice), the signal that develops below the 2D transition temperature is favored by $\pi$ incident photons, indicating its magnetic origin.}
\label{fig:pol-dep_LTHT}
\end{figure*}

\begin{figure*}[!ht]
\centering
\includegraphics[width=0.8\textwidth]{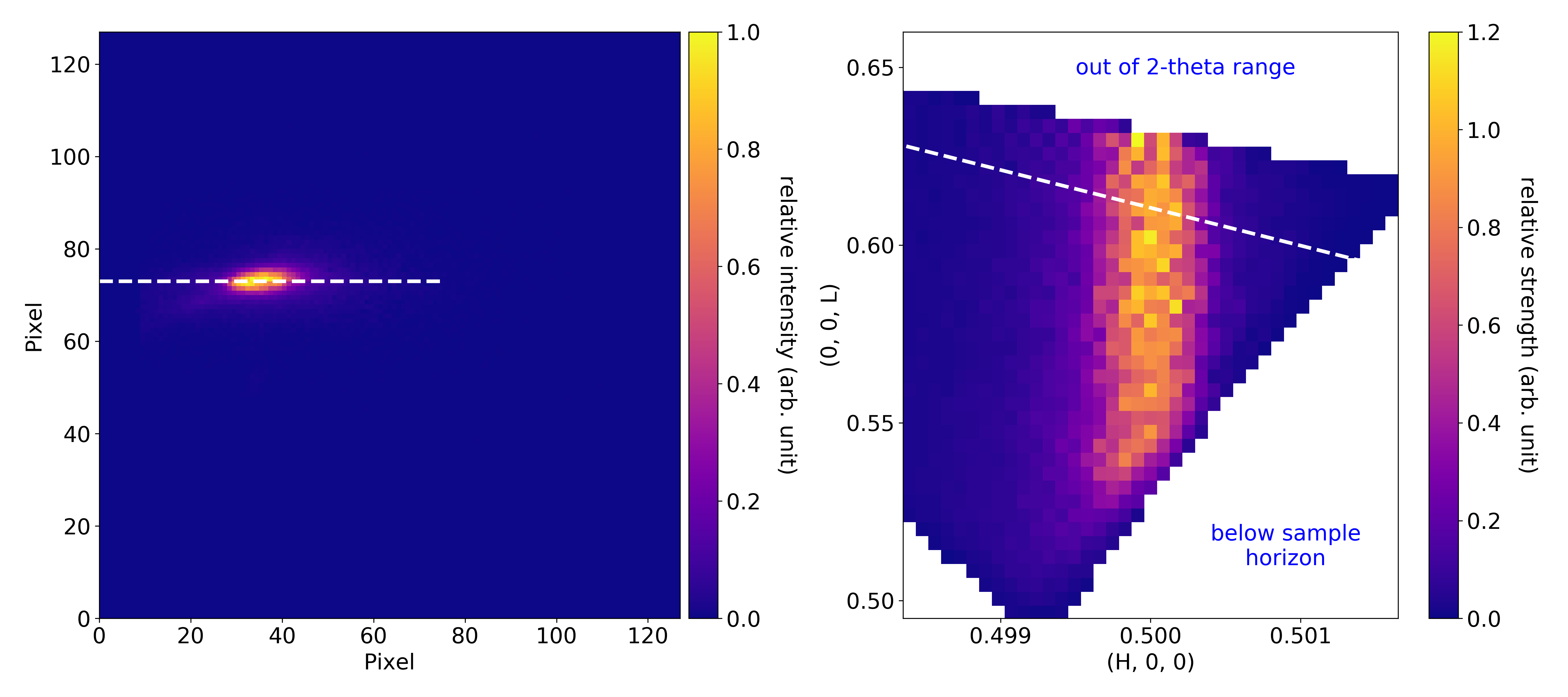}
\caption{Rod-like magnetic RXD signal at $\mathbf{q} = (0.5, 0, L)$. Left: a raw MCP intensity snap from the scan. The square-shaped MCP detector is covered by a round mask with a diameter of 2.5 cm; Right: Reconstructed signal in $(H,~0,~L)$ plane. The white dashed line on the left corresponds to the one on the right.}
\label{fig:M-LR_H0L-plane}
\end{figure*}

\begin{figure*}[!ht]
\centering
\includegraphics[width=0.8\textwidth]{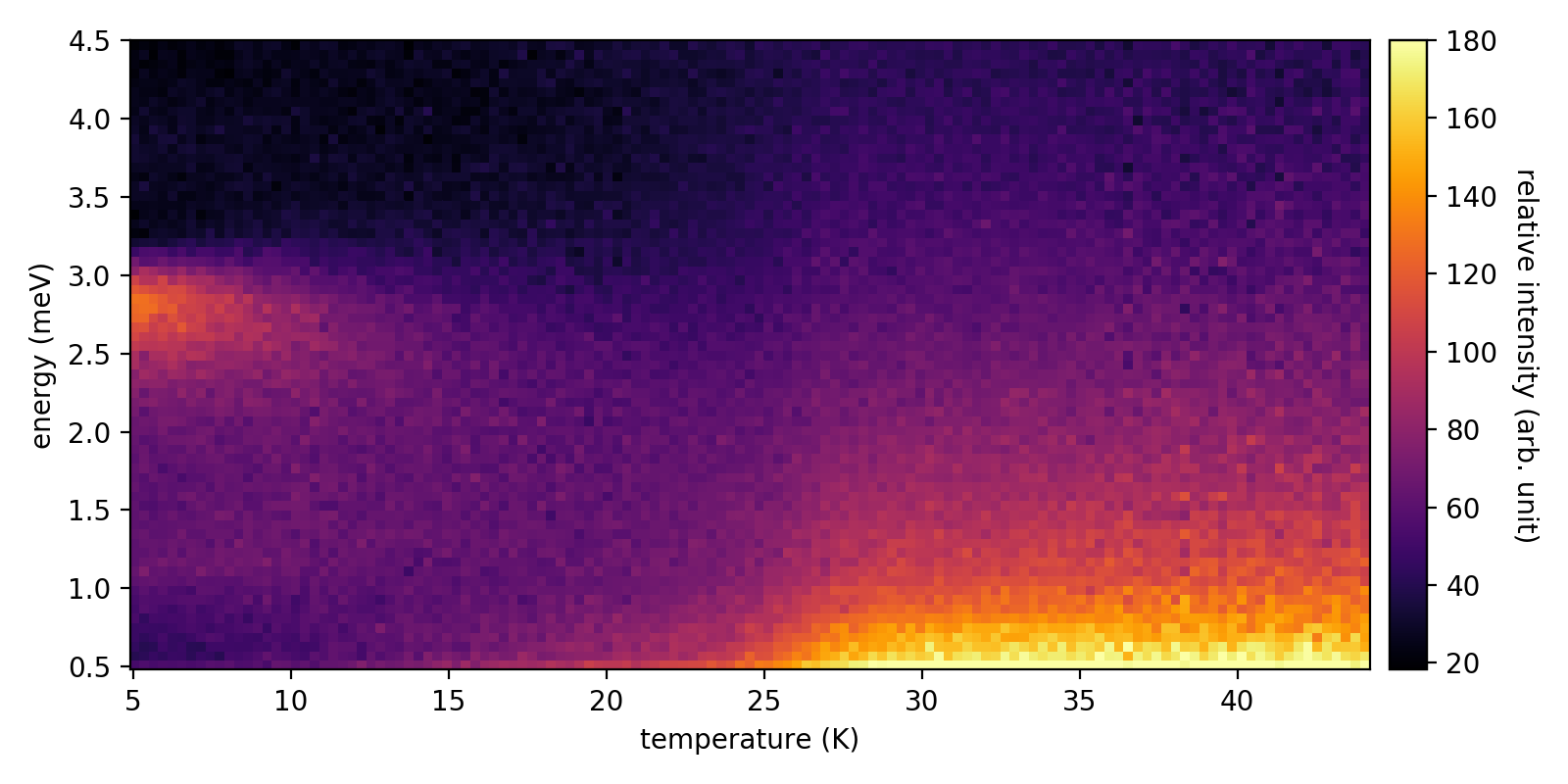}
\caption{INS intensities integrated over a large $\mathbf{q}$ range ($H \in [-0.9,~1.5],~K \in [-0.9,~0.9]$, all $L$ in available data).}
\label{fig:int_BZ_6meV_c2w_rebinned}
\end{figure*}

\begin{figure*}[!ht]
\centering
\includegraphics[width=0.8\textwidth]{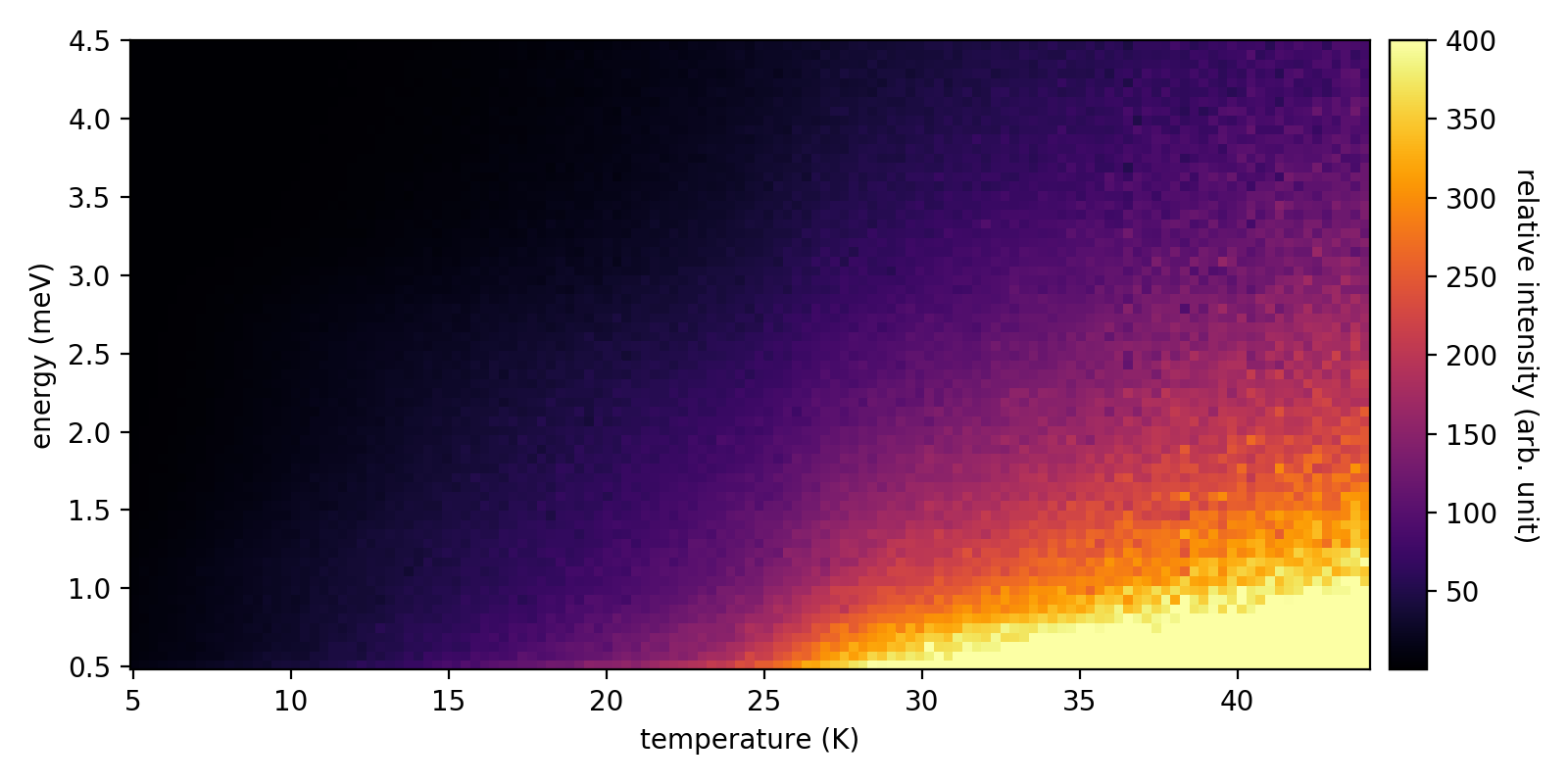}
\caption{Integrated INS intensities multiplied by the Bose-Einstein distribution function and energy transfer.}
\label{fig:dU_6meV}
\end{figure*}

\begin{figure*}[!ht]
\centering
\includegraphics[width=1\textwidth]{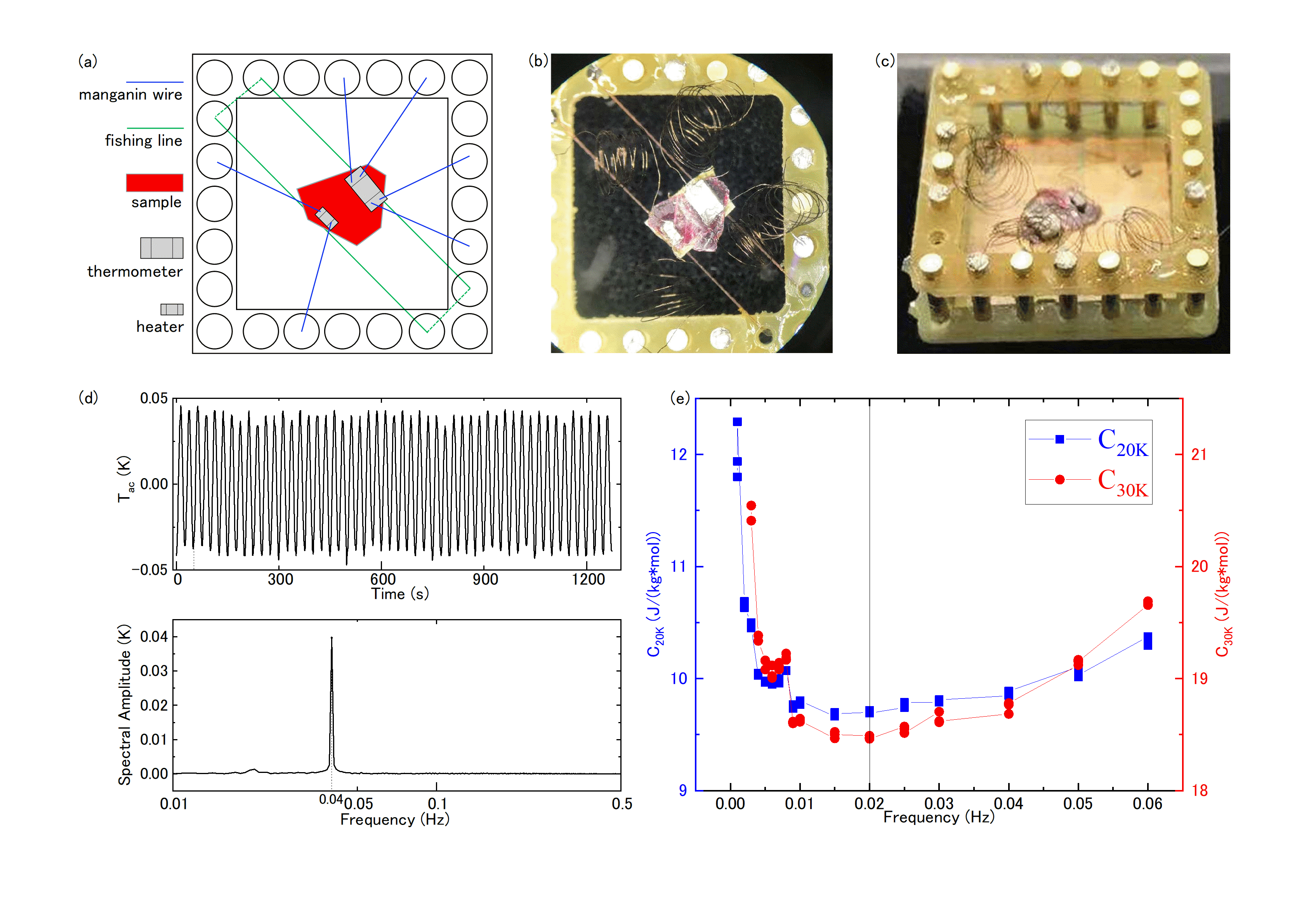}
\caption{(a-b) Schematic diagram and a photograph of the device. The sample, over a piece of insulating paper, was glued with GE varnish on two tightened parallel fishing lines attached to the sample holder. Spiral manganin wires were attached on the electrodes on the device with silver paint. (c) Photograph of the device for re-calibration of thermometer. After cutting the fish wire, we fixed the sample on a solid sample holder and plugged the original sample holder to the new solid one with all contacts unchanged. (d) Representative data sets from one sampling (upper) and its FFT result (bottom). The location of the frequency spectrum peak is exactly the heating power frequency, which is twice of the heating excitation current frequency. (e) Data of frequency scan at both 20 K and 30K. The appropriate heating excitation current frequency 0.02 Hz is marked by the solid vertical line. }
\label{fig:specific_heat_1}
\end{figure*}

Figure \ref{fig:NMR_SM}(a) displays our $^{23}$Na NMR spectra, measured under a low field of 0.75~T applied along the $\mathbf{a}^*$ direction, at selected temperatures. The NMR echo intensity is multiplied by temperature to correct for the total spectral weight, which scales with 1/$T$ as a results of nuclear magnetization. When cooled from 60~K down to 20~K, changes of the NMR lineshape are seen as a results of electronic magnetism. The total spectral weight, on the other hand, does not seem to change much when cooled from 60~K to 31~K. From 31~K down to 27~K, a reduction of the total spectral weight is clearly seen. This is further demonstrated by the integrated spectral weight as a function of temperature in Fig. \ref{fig:NMR_SM}(b). A clear decrease of the spectral weight is demonstrated when cooled below 31~K (labeled as $T^*$), until the spectral weight reaches the minimum at 27~K ($T_\mathrm{N}$). Below $T_N$, the spectral weight recovers, but still remains about 25$\%$ smaller than that at $T^*$.

For a magnetic system, the loss of the NMR spectral weight usually occurs close to a magnetic transition, so that part of the NMR signal is either out of the NMR frequency window due to magnetic inhomogeneity, or wiped out due to very slow dynamics. We believe that the latter scenario explains the decrease of the spectral weight at temperature below $T^*$, where the system develops quasi-static 2D magnetic ordering with long correlation lengths, as also revealed by the enhancement of $1/T_1$ below $T^*$ [Fig. 3(e) in the main text]. Below $T_\mathrm{N}$, the total spectra weight is partly recovered from the dip at $T_\mathrm{N}$, consistent with the 3D magnetic ordering. However, the $25\%$ loss persists even far below $T_\mathrm{N}$, which indicates that about one-quarter of the sample still maintains the cause for the loss. Such a behavior is not expected for the 3D ordering, but rather suggestive of the quasi-static ordering. Indeed, the latter is consistent with the 2D ordering observed by neutron diffraction: the ordering below $T^*$ features an $L$-independent diffraction pattern (see Fig. 2(a)), which suggests short-range ordering between the honeycomb layers. The relative volume of the 2D and 3D ordering, about 1:3 revealed by NMR, is also consistent with the neutron diffraction data concerning the ratio between the integrated intensities of the $L$-independent and the integer-$L$ components.



\section{Supplemental Figures}

\end{document}